\documentclass[utf8]{FrontiersinHarvard} 
\usepackage{url,hyperref,lineno,microtype,subcaption}
\usepackage[onehalfspacing]{setspace}
\usepackage{graphicx}
\usepackage{txfonts}
\usepackage[symbol]{footmisc}
\graphicspath{{./figures_jpg/}}
\def\firstAuthorLast{Kansabanik {et~al.}} 
\def\Authors{Devojyoti Kansabanik$^{1,2,*}$, Marcel Gouws$^{3}$, Deepan Patra$^{4}$, Angelos Vourlidas$^{2}$, Pieter Kotz\'{e}$^{5,3}$, Divya Oberoi$^{4}$, Shaheda Begum Shaik$^{6,7}$, Sarah Buchner$^{3}$, Fernando Camilo$^{3}$}
\def\Address{$^{1}$ NASA Jack Eddy Fellow, University Corporation for Atmospheric Research, Boulder, CO, USA \\
$^{2}$The Johns Hopkins University Applied Physics Laboratory, 11001 Johns Hopkins Rd, Laurel, MD, USA, 20723\\
$^{3}$South African Radio Astronomy Observatory, Liesbeek House, River Park, Gloucester Road, Mowbray, Cape Town, 7700, South Africa\\
$^{4}$National Centre for Radio Astrophysics, Tata Institute of Fundamental Research, S. P. Pune University Campus, Pune, India\\
$^{5}$ National Radio Astronomy Observatory, 520 Edgemont Road, Charlottesville, VA 22903 USA\\
$^{6}$George Mason University, Fairfax, VA 22030, USA\\
$^{7}$ U.S. Naval Research Laboratory, Washington, DC 20375, USA}

\def\corrAuthor{Devojyoti Kansabanik}
\def\corrEmail{devojyoti96@gmail.com, dkansabanik@ucar.edu}

\begin{document}
%\onecolumn
\firstpage{1}

\title[MeerKAT Solar Observation]{Solar Observation with MeerKAT: Demonstration of Technical Readiness and Initial Science Highlights} 

\makeatletter
\def\@extraAuth{}
\makeatother
\author[\firstAuthorLast ]{\Authors} %This field will be automatically populated
\address{} %This field will be automatically populated
\correspondance{} %This field will be automatically populated

\maketitle
\begin{abstract}
Solar radio emissions offer unique diagnostic insights into the solar corona. However, their dynamic and multiscale nature, along with several orders of magnitude variations in intensity, pose significant observational challenges. To date, at gigahertz frequencies, MeerKAT stands out globally with high potential of producing high-fidelity, spectroscopic snapshot images of the Sun, enabled by its dense core, high sensitivity, and broad frequency coverage. Yet, as a telescope originally designed for observing faint galactic and extragalactic sources, observing the Sun at the boresight of the telescope requires customized observing strategies and calibration methods. This work demonstrates the technical readiness of MeerKAT for solar observations at the boresight of the telescope in the UHF (580–1015 MHz) and L-band (900–1670 MHz) frequency ranges, including optimized modes, a dedicated calibration scheme, and a tailored, entirely automated calibration and imaging pipeline. The quality of solar images is validated through morphological comparisons with the solar images at other wavelengths. Several unique early science results showcase the potential of this new capability of MeerKAT. Once fully commissioned and operational, this will unlock novel solar studies, significantly expand the scientific portfolio of MeerKAT, and lay the groundwork for solar observations with the mid-frequency telescope of the upcoming Square Kilometre Array Observatory, for which MeerKAT serves as a precursor.
\end{abstract}

\section{Introduction}\label{sec:intro}
The solar atmosphere consists of hot, magnetized plasma, with thermal and non-thermal electrons producing radio emissions across a broad frequency range (kHz to GHz) through processes like thermal bremsstrahlung, plasma emission, gyro-resonance, gyro-synchrotron, and electron-cyclotron maser emission. Solar radio emissions have been studied extensively from a few kHz to hundreds of GHz \citep[e.g.][]{Pick2008, Gary2023_review}. Most observations rely on spectrograms, which provide spectrotemporal data but lack spatial information. Solar imaging at radio wavelengths has been carried out for decades by a small number of dedicated interferometers, including the Nan\c{c}ay Radio Heliograph \citep[NRH;][]{bonmartin1983} and Gauribidanur Radio Heliographs \citep[GRAPH;][]{sundaram2004}, both still operational, and the Nobeyama Radioheliograph \citep[NoRH;][]{Nakajima1994}, which is no longer active. This has changed with new-generation radio interferometers such as the Murchison Widefield Array \citep[MWA;][]{lonsdale2009,tingay2013}, LOw Frequency ARray \citep[LOFAR;][]{lofar2013}, upgraded Giant Metrewave Radio Telescope \citep[uGMRT;][]{Gupta_2017}, Jansky Very Large Array\citep[JVLA;][]{VLA2009}, the Expanded Owens Valley Solar Array \citep[EOVSA;][]{Gary2012}, and Atacama large millimeter-submillimeter array \citep[ALMA;][]{Bastian2022_alma,Shimojo2024}. Although not all of them focus solely on solar observations, they have been instrumental in advancing our understanding of solar physics.

MeerKAT, a new-generation radio interferometric telescope in South Africa \citep{meerkar2016} and a precursor to the mid-frequency telescope of the upcoming Square Kilometre Array Observatory \citep[SKAO,][]{ska_concept,SKAO2021}, comprises 64 cryogenically cooled 13.5 m dishes with excellent sensitivity. It operates across UHF (580–1015 MHz), L (900–1670 MHz), and S (1750–3500 MHz) bands. Its dense core, 39 dishes within 1 km, and extended baselines up to 8 km provide superb surface brightness sensitivity and a well-sampled Fourier coverage, enabling high-fidelity spectroscopic snapshot imaging, even off-boresight \citep{Kansabanik_2024_meerkat}. These characteristics make MeerKAT an ideal instrument for studying the dynamic Sun at GHz frequencies, including active emissions, coronal mass ejection (CME) magnetometry \citep[e.g.,][]{Kansabanik_cme1,Kansabanik2024_cme2}, and faint transient detections. Although significant advances have been made at meter wavelengths with SKA-Low precursor (MWA) \citep[][and references therein]{Oberoi2023} and pathfinders (LOFAR, uGMRT) \citep[e.g.,][]{Magdalenic2020,Zhang2024,Mondal_2024_gmrt}, solar studies using precursor and pathfinders at mid-frequency of SKAO remain in early stages.

The first MeerKAT application to solar science was conducted by \citet{Kansabanik_2024_meerkat}, showcasing its potential for high-fidelity solar imaging. These observations placed the Sun in a sidelobe of the primary beam of the telescope, rather than at boresight, to sufficiently attenuate the intense solar emission and enable stable operation of the signal chain. However, the large angular size of the Sun and the chromaticity of the primary beam sidelobes posed significant challenges, necessitating attenuation that reduced sensitivity by a factor of $\sim 1000$. The complex, evolving structures of the Sun that span arcseconds (the smallest detectable scale is often set by coronal scattering \citep{Bastian1994}) to full-disk scales at GHz frequencies pose substantial calibration challenges. Successfully addressing these requires a two-step approach: (1) optimizing telescope configuration for solar observations and (2) calibrating data to match this new configuration.

This paper demonstrates the technical readiness of MeerKAT for solar observations with the Sun positioned at the boresight of the telescope, including a dedicated pipeline for calibrating these non-standard observations. The structure of the paper is as follows: Section \ref{sec:meersolar} details system optimization and observing procedures. Section \ref{sec:solar_motion} addresses the effects of the motion of the Sun on the sky, followed by a description of the data processing pipeline, including calibration, imaging, and mitigation strategies for the effects of non-sidereal solar motions in Section \ref{sec:calibration}. Section \ref{sec:verify_system} verifies system performance and demonstrates the technical readiness. We highlight several preliminary interesting science results in Section \ref{sec:early_science_results}. Section \ref{sec:conclusion} concludes the paper with a discussion about future works.

\section{Configuring MeerKAT for Solar Observations}\label{sec:meersolar}

\subsection{Challenges in Observing the Sun with MeerKAT}\label{subsec:solar_spectrum}
The Sun is the source with the highest flux density at GHz frequencies and has a large angular size ($\geq32'$). The full-width half-maximum of the primary beam of MeerKAT is about 2$^\circ$ and 1.2$^\circ$ at the central frequency of UHF- and L-band, respectively \citep{deVilliers2023}. Hence, the Sun fills a significant portion of the primary beam of MeerKAT at UHF and L-band. The Sun fills the entire beam at the higher frequency part of S-bands.  This results in a significantly higher beam-integrated power compared to typical astronomical sources, necessitating strong attenuation to ensure that the signal chain of the telescope remains within its optimal regime. However, this creates a calibration challenge. The quiet Sun has a flux density of a few hundred Solar Flux Unit (sfu) (1 sfu = $10^4$ Jansky (Jy)), increasing with frequency due to its thermal nature and reaching several thousand sfu due to non-thermal emissions during the presence of solar activity. In contrast, all bright astronomical calibrators \citep{Perley_2017} as well as the A-team sources \citep[Cassiopeia A, Cygnus A, Taurus A, Virgo A;][]{Gasperin2020} have negative spectral indices, with flux densities that decrease with frequency. As a result, they cannot be observed with the same attenuation settings as the Sun. This makes traditional flux density calibration using astronomical calibrators infeasible for solar observations with MeerKAT, necessitating an independent method to characterize the spectral response of the attenuators. To prepare MeerKAT for solar observations, we have performed several engineering tests between late 2022 and late 2023 under the project ID: EXT-20221114-PK-01. 

\subsection{Description of Signal Power Management at MeerKAT for Solar Observations}\label{subsec:signal_manage}
In general, the low-noise amplifier (LNA) -- the first component in the signal chain of a radio telescope -- is designed to have a linear response over a wide dynamic range, enabling it to accommodate strong signals such as those from the Sun. However, downstream sub-systems, including those of MeerKAT, have more limited dynamic ranges, and the default configuration is optimized for observing faint astronomical sources. To manage strong solar signals, MeerKAT employs attenuators within the Radio Frequency Conditioning Unit (RFCU). This is a room-temperature subsystem located just before the analog-to-digital converter (ADC) in the signal chain. These attenuators offer 0–63 dB attenuation in 1 dB steps. To calibrate this attenuation, we have used a built-in noise diode. The built-in noise diode injects noise with a temperature approximately equal to system temperature on cold sky, leading to an increase in power by approximately 3 dB. Hence, measuring the change corresponding to the power injected by the noise diode when using attenuators allows us to measure the effective attenuation.

We estimated the additional signal attenuation required for MeerKAT solar observations to maintain the input power to the ADCs near the nominal level, based on source flux density and the band-averaged System Equivalent Flux Density (SEFD). Under cold-sky conditions, MeerKAT sets attenuation to align ADC input power to the nominal $-29$\,dBFS (dBFS: Decibels relative to full scale; 0 dBFS is the digital maximum), with $-13$\,dBFS as the upper operational limit. Using a band-averaged SEFD (\href{https://skaafrica.atlassian.net/wiki/spaces/ESDKB/pages/277315585/MeerKAT+specifications\#System-Equivalent-Flux-Density-(SEFD)}{MeerKAT SEFD specifications}) and typical mean solar flux densities of $100$\,sfu (UHF) and $200$\,sfu (L-band), the required additional attenuation is given by
\begin{equation}
    S_\mathrm{att,dB}(S_\mathrm{sun}) = 10 \log_{10} \left( \frac{S_\mathrm{sun} + \mathrm{SEFD}}{\mathrm{SEFD}} \right),
    \label{eq:estimate_att}
\end{equation}
yielding $\sim$32\,dB and $\sim$35\,dB for UHF- and L-bands, respectively. These estimates were validated with solar test observations on 11 November 2022 (UHF) and 11 January 2023 (L-band). 

These estimates are based on quiet solar flux density and are designed to set the ADC power to the {\it nominal power level}, which is the minimum input power level required for optimal operation of the ADC. Hence, this value of $S_\mathrm{att}$ provides the maximum possible headroom needed to accommodate the increased flux density during solar flares.  While the estimates of solar flux density mentioned above are representative, the disc-integrated quiet Sun solar flux density can vary on timescales of a day. To account for this time variability, we developed a flexible system that automatically obtains the quiet solar flux density ($S_\mathrm{sun}$) from the previous day measurements from the Learmonth Solar Radio Observatory; alternatively, this information can also be provided by the user. This system estimates and applies the required additional attenuation ($S_\mathrm{att,dB}$) following Equation \ref{eq:estimate_att}.

\subsection{Characterization of the Attenuators}\label{subsec:att_character}
Since attenuators introduce an additional element into the signal path, it is essential to assess their impact on spectral properties, as well as their influence on visibility amplitudes and phases. To achieve this, we analyze the variations in visibility amplitudes and phases for various attenuation levels, while ensuring that the ADC power stays within its optimal operating range. The latter is needed to ensure that the signal-to-noise ratio (SNR) of visibility does not change significantly.

\begin{figure*}[!htbp]
    \centering
    \includegraphics[trim={0.5cm 0cm 0cm 0.5cm},clip,width=0.46\linewidth]{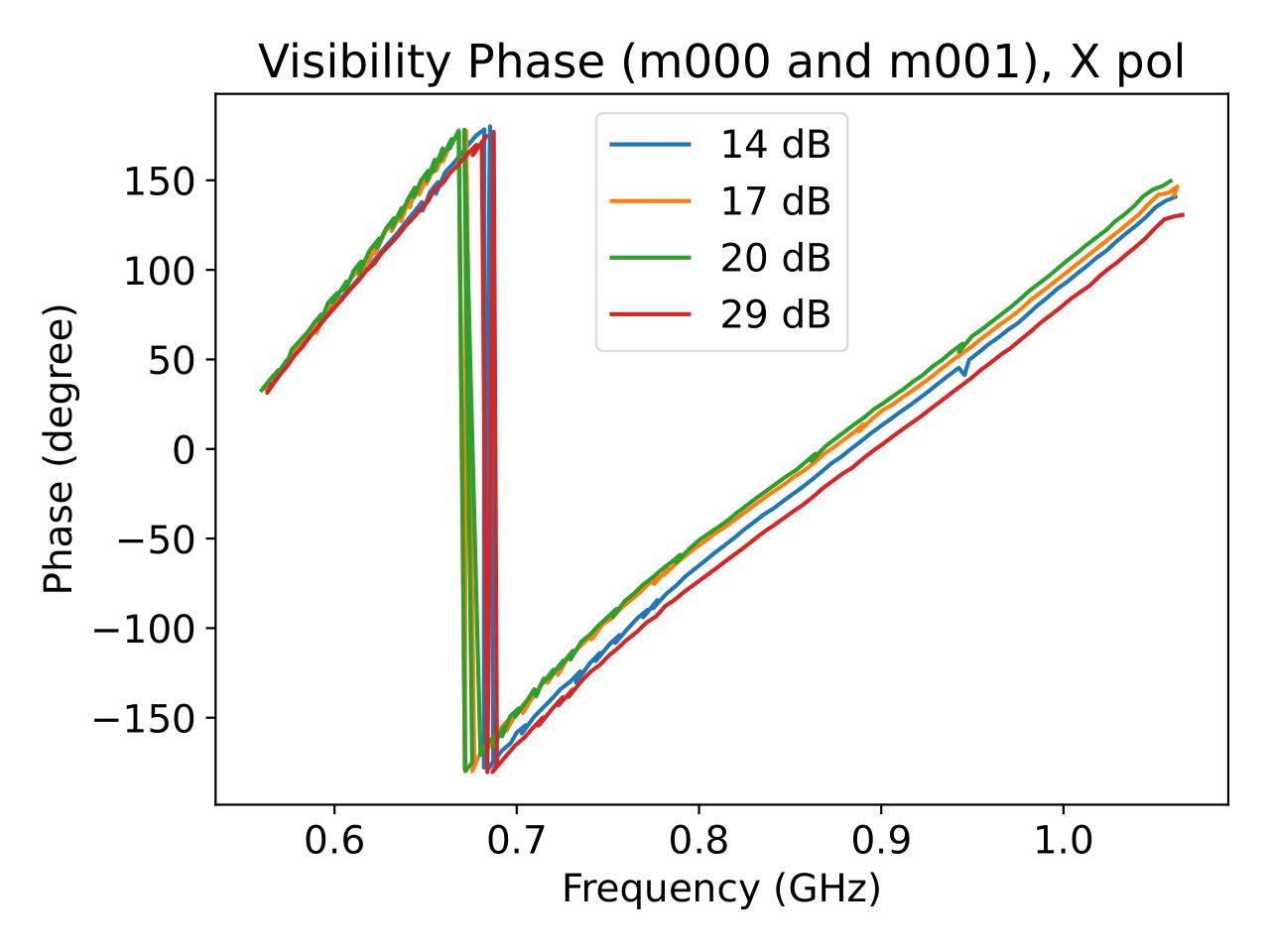}
    \includegraphics[trim={1.1cm 0cm 0cm 0.5cm},clip,width=0.44\linewidth]{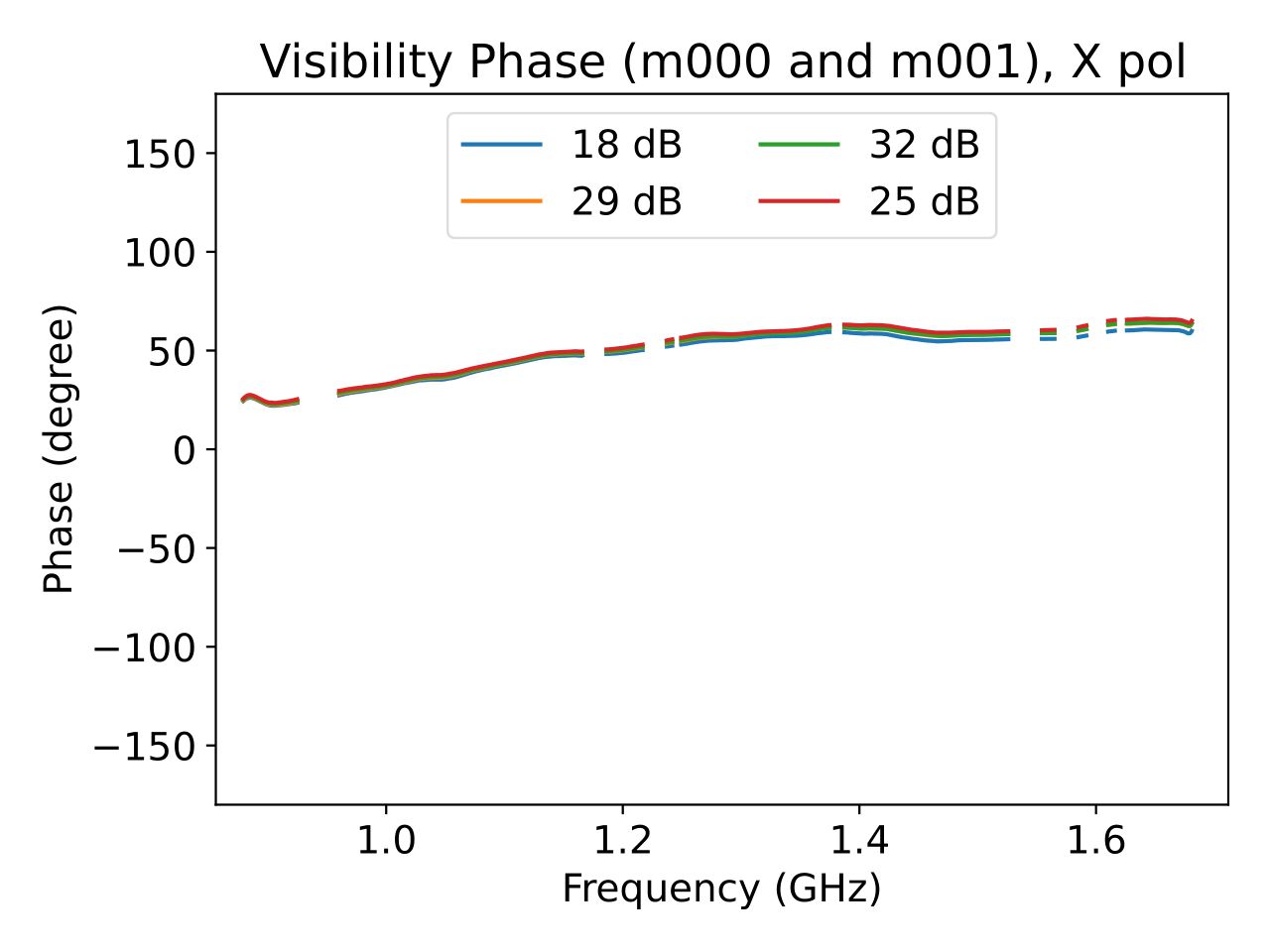}\\
    \includegraphics[trim={0.5cm 0cm 0cm 0.5cm},clip,width=0.46\linewidth]{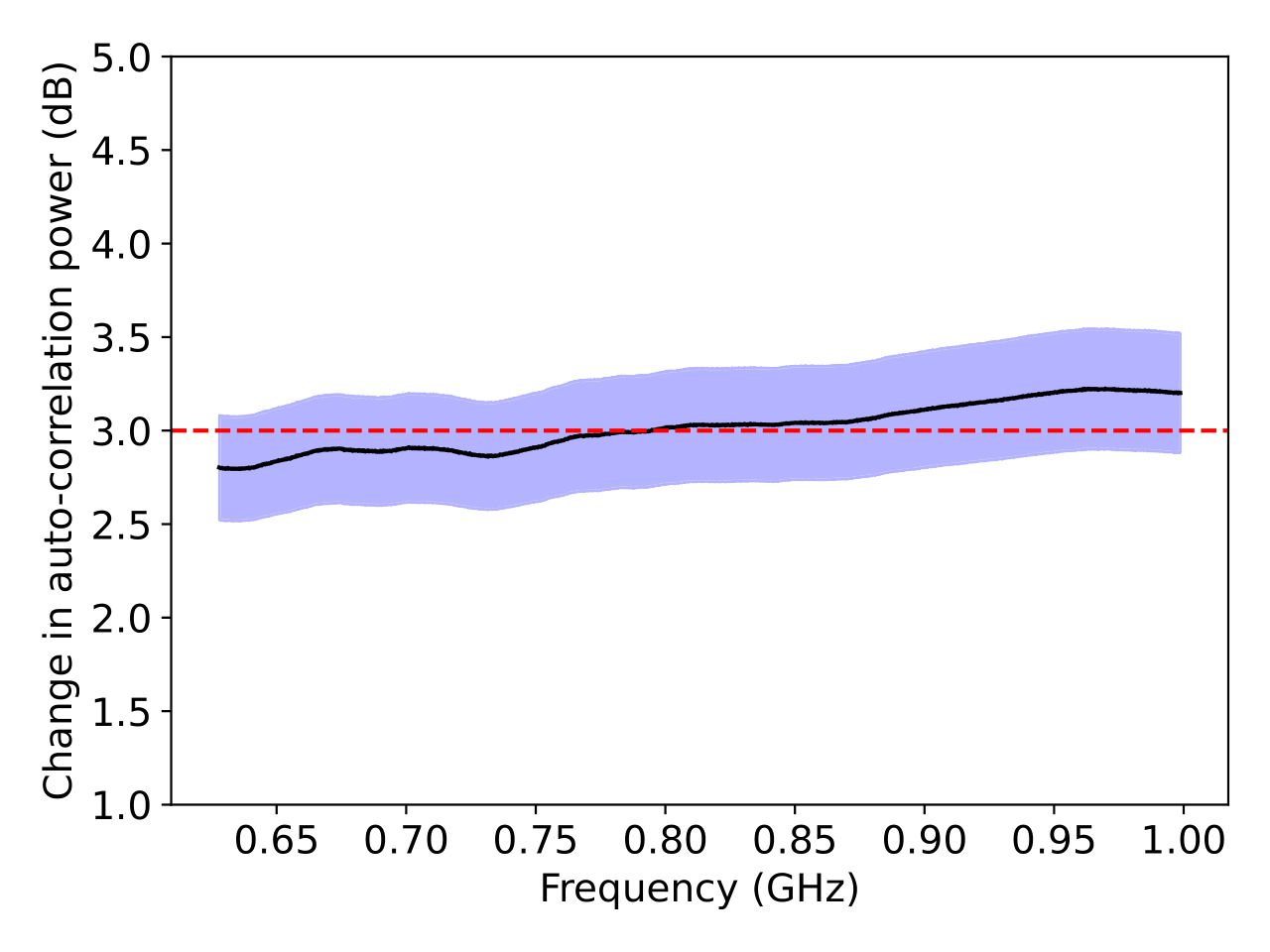}\includegraphics[trim={1cm 0cm 0cm 0.5cm},clip,width=0.44\linewidth]{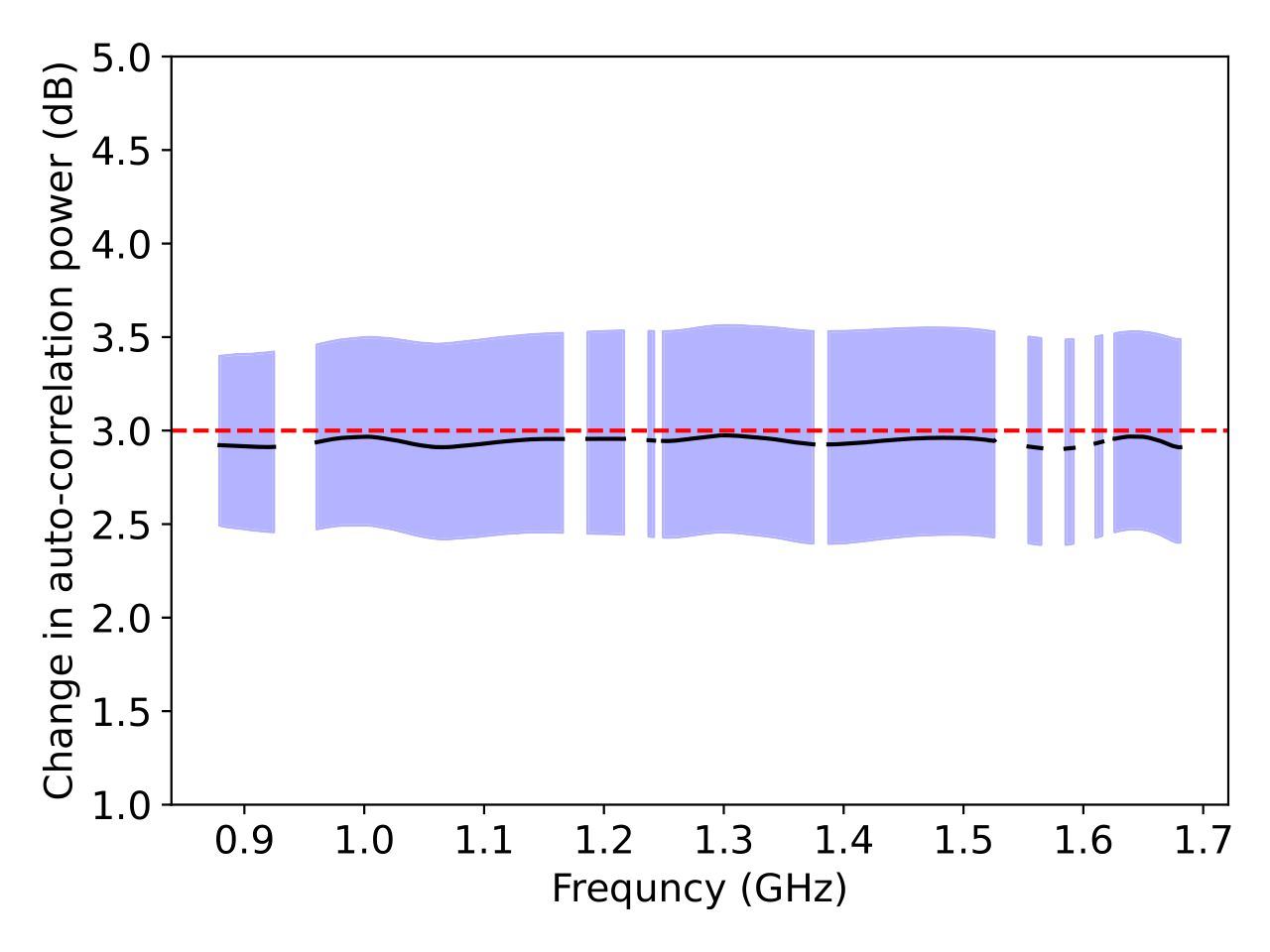}
    \caption{{\it Top panels:} Visibility phases for X-polarization for baseline between antennas m000 and m001 (baseline length of 37 meters) for different values of $S_\mathrm{att,dB}$. {\it Bottom panels:} Spectral variation of auto-correlation power for 3\,dB changes in $S_\mathrm{att,dB}$ between 29 and 32\,dB. The solid black line shows the power level change averaged over all antennas, while the blue shaded region indicates the standard deviation about the mean. The red dashed line marks the expected 3\,dB attenuation. The left and right panels represent UHF and L-bands, respectively. Y-polarization shows a similar behavior. }
    \label{fig:att_char}
\end{figure*}

\subsubsection{Phase and Amplitude Distortion}\label{subsec:amp_phase}
To understand the phase response of attenuators, we compared visibility phases on the baseline between antennas m000 and m001 (baseline length of 37 meters) across multiple scans on the Sun with varying attenuation levels ($S_\mathrm{att,dB}$). As shown in the top panels of Figure \ref{fig:att_char}, the phases for X-polarization (similar for Y-polarization) remain consistent within $\pm5$ degrees. Although the attenuation change ($S_\mathrm{att,dB}$) is similar across antennas, these changes happen with respect to different initial attenuation settings. Initial attenuation values are different across antennas needed to arrive at {\it nominal power level} on the cold sky. As a result, the phase shifts introduced in different antennas may differ. The small differences in phases seen in the top panel of Figure \ref{fig:att_char} may arise from this or from the variations in solar emission itself. However, these small changes can conveniently be calibrated out during self-calibration. 

To assess the spectral behavior of the amplitude response of attenuators, we analyzed the auto-correlation power of all antennas in both UHF and L-bands, with $S_\mathrm{att,dB}$ varied in 3\,dB steps. The bottom panels of Figure \ref{fig:att_char} show that while power changes closely match the expected 3\,dB increments in L-band, in the UHF-band changes show a small frequency-dependent variation. Self-calibration can correct for antenna-to-antenna phase variations due to an additional attenuator. However, due to reasons mentioned in Section \ref{subsec:solar_spectrum}, the spectral response of the amplitude of the attenuators can not be calibrated using astronomical sources. Hence, we calibrated this absolute flux scaling and spectral response of attenuators using built-in noise diodes. 

\subsubsection{Antenna-to-antenna Variation}\label{subsec:ant_variation}
\begin{figure}[!htbp]
    \centering
    \includegraphics[trim={0.5cm 0.5cm 1cm 1cm},clip,width=\linewidth]{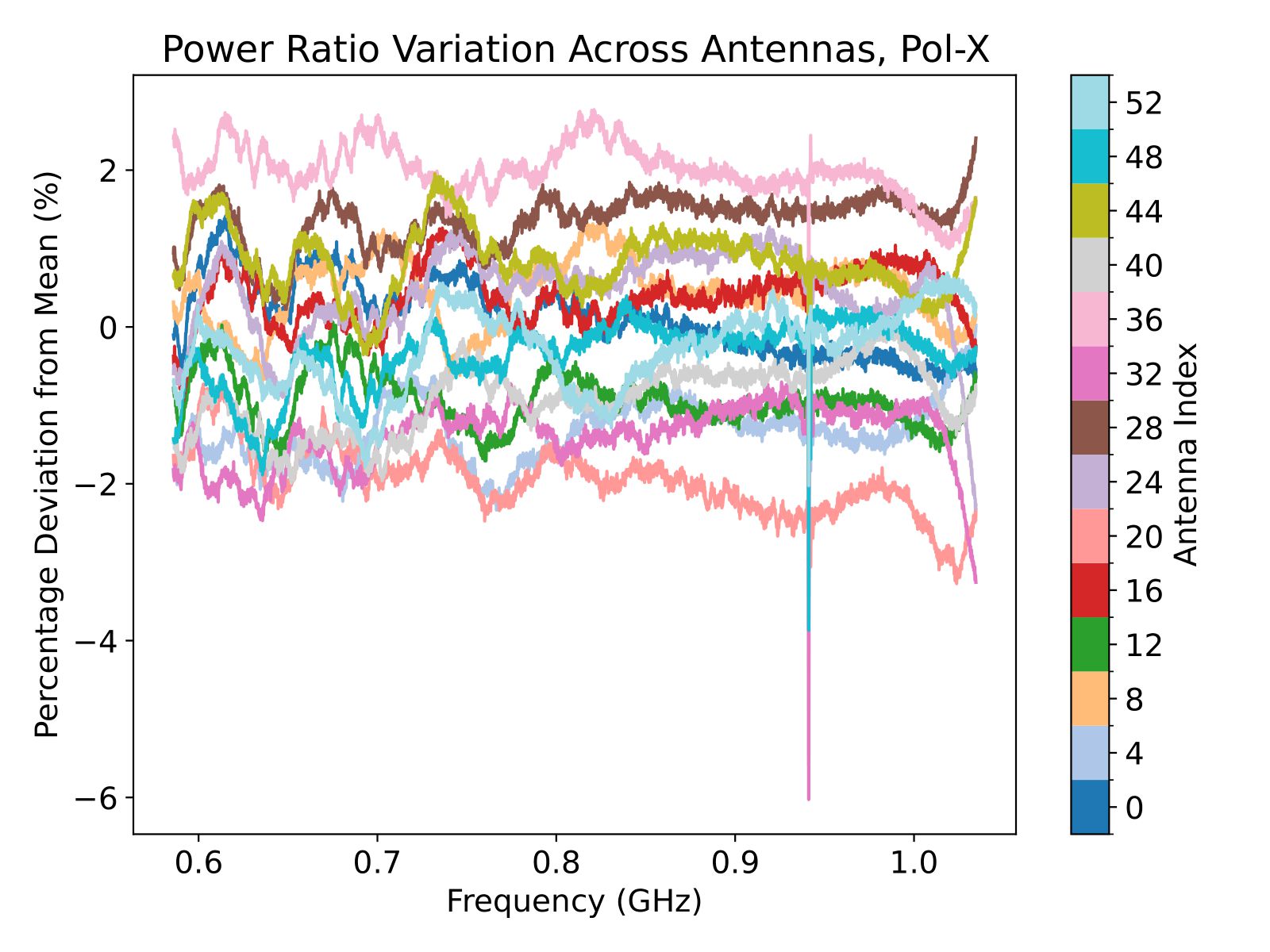}
    \caption{The relative variation in power level across different antennas, resulting from a 3 dB change in $S_\mathrm{att,dB}$, is shown as a percentage deviation from the mean change for X-polarization. The same is true for Y-polarization.}
    \label{fig:ant_variations}
\end{figure}

While $S_\mathrm{att,dB}$ is used uniformly in all antennas, actual adjustments in power level can vary between antennas due to the distinct physical nature of their attenuators. Figure \ref{fig:ant_variations} illustrates the percentage deviation from the mean spectral change across all antennas to highlight the extent of these variations. The antenna-to-antenna attenuation variations are sufficiently small, within $\pm2\%$. Hence, one can average over multiple antennas to build up the SNR for estimating the attenuation value using the noise diodes. 

\subsection{Observing Strategy With Noise Diode for Flux Density Calibration}\label{subsec:noise_diode}
As discussed in Sections \ref{subsec:solar_spectrum} and \ref{subsec:att_character}, standard calibrators including the A-team sources cannot be observed with the same attenuation as used for the Sun, and the applied $S_\mathrm{att,dB}$ exhibits a non-flat spectral response that must be characterized for absolute flux calibration. This is achieved using the built-in noise diode. Located after the LNA and before the attenuator in the signal chain, they inject a power of known strength, similar to the system temperature, $T_\mathrm{sys}$, in both UHF and L-band receivers. During calibrator scans without the additional attenuation, the noise diode-induced power change is $d_\mathrm{cal}(\nu) = P_\mathrm{cal,on}(\nu) - P_\mathrm{cal,off}(\nu)$, where $\nu$ is the frequency and the subscripts on and off refer to the observed power with the noise diode switched on and off, respectively. During solar scans with additional attenuation, it is $d_\mathrm{sun}(\nu) = P_\mathrm{sun,on}(\nu) - P_\mathrm{sun,off}(\nu)$. Hence, the ratio $d_\mathrm{sun}(\nu)/d_\mathrm{cal}(\nu)$ captures the spectral variation in the attenuation, enabling calibration of the attenuator response and determination of the absolute flux density of the solar observations. The presence of additional attenuation $S_\mathrm{att,dB}$ reduces the effective noise diode power during solar observations, making it difficult to achieve sufficient SNR, even when the noise power is similar to the $T_\mathrm{sys}$. As discussed in Section \ref{subsec:fluxcal_noisediode}, this necessitates longer integration over time and/or frequency. The value of the noise diode being used for MeerKAT is close to the minimum needed for calibrating the attenuator response for solar observations. Reducing them below their current levels will require too long integrations, which is likely to make their use for solar flux density calibration impractical. 

\subsection{Standard Observing Procedure of Solar Observation with MeerKAT}\label{sec:sop}
In preparation for solar observations with MeerKAT, we conducted tests in the engineering mode and arrived at the following observing procedure for routine solar observations:
\begin{enumerate}
    \item \textbf{Flux calibrator scan:} Observe a standard MeerKAT flux/bandpass calibrator (e.g., J1939-6342 or J0408-6545 ({\href{https://skaafrica.atlassian.net/wiki/spaces/ESDKB/pages/1481408634/Flux+and+bandpass+calibration}{MeerKAT flux and bandpass calibrators}}) with nominal attenuation.
    \item \textbf{Noise diode calibration on calibrator:} Perform a 3–5 minute scan on the same calibrator, switching the noise diode on and off for every successive correlator integration, so that consecutive data records alternate between noise-on and noise-off states. (Currently implemented via engineering mode).
    \item \textbf{Phase and polarization calibrator:} Observe suitable phase and polarization calibrators with nominal attenuation settings.
    \item \textbf{Point to the Sun:} Slew to the Sun with nominal attenuation still active. Switching on solar attenuation before the Sun is in the primary beam may cause system issues.
    \item \textbf{Enable solar attenuation:} Once on-source, activate the additional solar attenuation ($S_\mathrm{att,dB}$).
    \item \textbf{Solar scan with noise diode:} Conduct solar scans with the noise diode toggled on alternate correlator dumps. It is recommended to limit the scan duration to 30 minutes for phase calibrator observations. 
    \item \textbf{Disable attenuation:} After the solar scan, disable $S_\mathrm{att,dB}$ before slewing away from the Sun.
    \item \textbf{Post-scan calibrator:} Re-observe the phase calibrator with standard attenuation.
    \item \textbf{Repeat cycle:} Repeat steps 4-7 for the remaining observing time.
\end{enumerate}
We note that once the functionality required for inserting appropriate attenuation in the signal path and for toggling the noise diode on alternate correlator dumps is implemented in the MeerKAT Observation Planning Tool, all essential requirements for enabling a solar observing mode will have been met.

\subsection{Minimum Pointing Distance of Calibrators from the Sun}\label{subsec:calibrator_pointing}
When the telescope points near the Sun, the system temperature can increase significantly. For the MeerKAT beam, this minimum angular distance, $D_\mathrm{sun,min}$, is approximately $7^\circ$ in UHF and $4.5^\circ$ in L-bands (\href{https://archive-gw-1.kat.ac.za/public/meerkat/Solar-avoidance-radius.jpg}{MeerKAT Solar Avoidance Zone}). Therefore, calibrators or astronomical targets should be observed at angular distances greater than $D_\mathrm{sun,min}$. Additionally, solar wind turbulence can introduce phase errors and scatter broadening in calibrator observations near the Sun. The phase error due to such turbulence can be estimated using (\href{https://library.nrao.edu/public/memos/vla/test/VLAT_236.jpg}{Butler 2005, NRAO}):
\begin{equation}
    R_\mathrm{degree} \sim \left(\frac{7\ \lambda_\mathrm{cm}\ B_\mathrm{km}^{0.29}}{\phi_\mathrm{degree}}\right)^{0.71},
\end{equation}
where $R_\mathrm{degree}$ is the minimum angular distance from the Sun (in degrees), $\lambda_\mathrm{cm}$ the wavelength (in cm), $B_\mathrm{km}$ the baseline (in km), and $\phi_\mathrm{degree}$ the allowable phase error (in degrees). For MeerKAT, assuming $\phi_\mathrm{degree} = 10^\circ$, this yields $R_\mathrm{degree} \sim 15^\circ$ (UHF) and $10^\circ$ (L-band), as detailed in Table \ref{table1}. 
\begin{table*}[!htbp]
    \centering
    \begin{tabular}{|l||c|c|l|l|l|l|} \hline 
          Band&$\phi_\mathrm{1,degree}$&  $R_\mathrm{1,degree}$&  $\phi_\mathrm{5,degree}$&$R_\mathrm{5,degree}$& $\phi_\mathrm{10,degree}$&$R_\mathrm{10,degree}$\\ \hline \hline
          UHF-band (37 cm)&  1.0& 79.0&  5.0&25.0& 10.0&15.0\\ \hline
        L-band (21 cm)& 1.0& 53.0&  5.0&16.0& 10.0&10.0 \\\hline
    \end{tabular}
    \caption{Minimum angular distance of the calibrators from the Sun for specified expected phase errors.}
    \label{table1}
\end{table*}

\section{Effects of Motion of the Sun}\label{sec:solar_motion}
The apparent motion of the Sun in the sky is unlike that of most astronomical sources. It is governed by two key components: the non-sidereal motion of the Sun on the sky, and the movement of solar features on the solar disc due to differential solar rotation. Both these effects must be considered when observing and analyzing solar data.

\subsection{Sidereal Motion}\label{subsec:sidereal_motion}
The Sun, being a non-sidereal source, its Equatorial coordinates (RA-Decl.) drift across the sky at an average rate of $\sim1^\circ$ per day ($\sim2.5^{\prime\prime}$ per minute), giving rise to a uniform shift in the coordinates of solar features. In radio interferometry, {\it delay-tracking} at the correlator compensates for geometric delays between received signals at different antennas as the source moves. While sidereal sources are tracked at a fixed equatorial coordinate, solar observations with a telescope require tracking the solar center. The correlator delay center is continuously updated at a kHz rate using a linear model, with model parameters -- delay and delay-rate -- estimated and refreshed every 5 seconds. This approach ensures accurate correlation and prevents decorrelation due to the solar apparent motion.

\subsection{Differential Rotation}\label{subsec:diff_rotation}
In addition to the non-sidereal motion, the Sun also exhibits differential rotation -- the rotation period at its equator is $\sim$25 days and that at polar regions is $\sim$34 days \citep{Mancuso2020}. This causes solar features at different latitudes to move in the plane of the sky at varying rates, with the maximum projected motion occurring near the solar disk center. The maximum differential motion is, $\theta_\mathrm{diff,rot} \approx 5.0^{\prime\prime}/\mathrm{hour}$. To limit smearing arising due to differential rotation, even after correcting for the overall non-sidereal motion of the Sun, integration times should not exceed $\sim$130 minutes in the UHF band and $\sim$50 minutes in the L-band. 

We note that differential solar rotation breaks the ``rigid-sky" assumption of radio interferometric imaging, making corrections in the visibility domain or during the imaging and deconvolution non-trivial. Current tools like Common Astronomy Software Application \citep[CASA;][]{CASA2022} and W-Stacking CLEAN \citep[WSClean;][]{Offringa2014} do not support such corrections. This limitation will become more critical for the SKAO, with its higher spatial resolution, where uncorrected differential rotation may smear fine-scale features even when integrating over short times. For example, maximum integration time should be less than $\sim$5 minutes at 1 GHz to avoid smearing due to differential rotation. A dedicated imaging algorithm to address this is under development and will be presented in a forthcoming publication.

\begin{figure*}[!htbp]
    \centering
    \includegraphics[width=\linewidth]{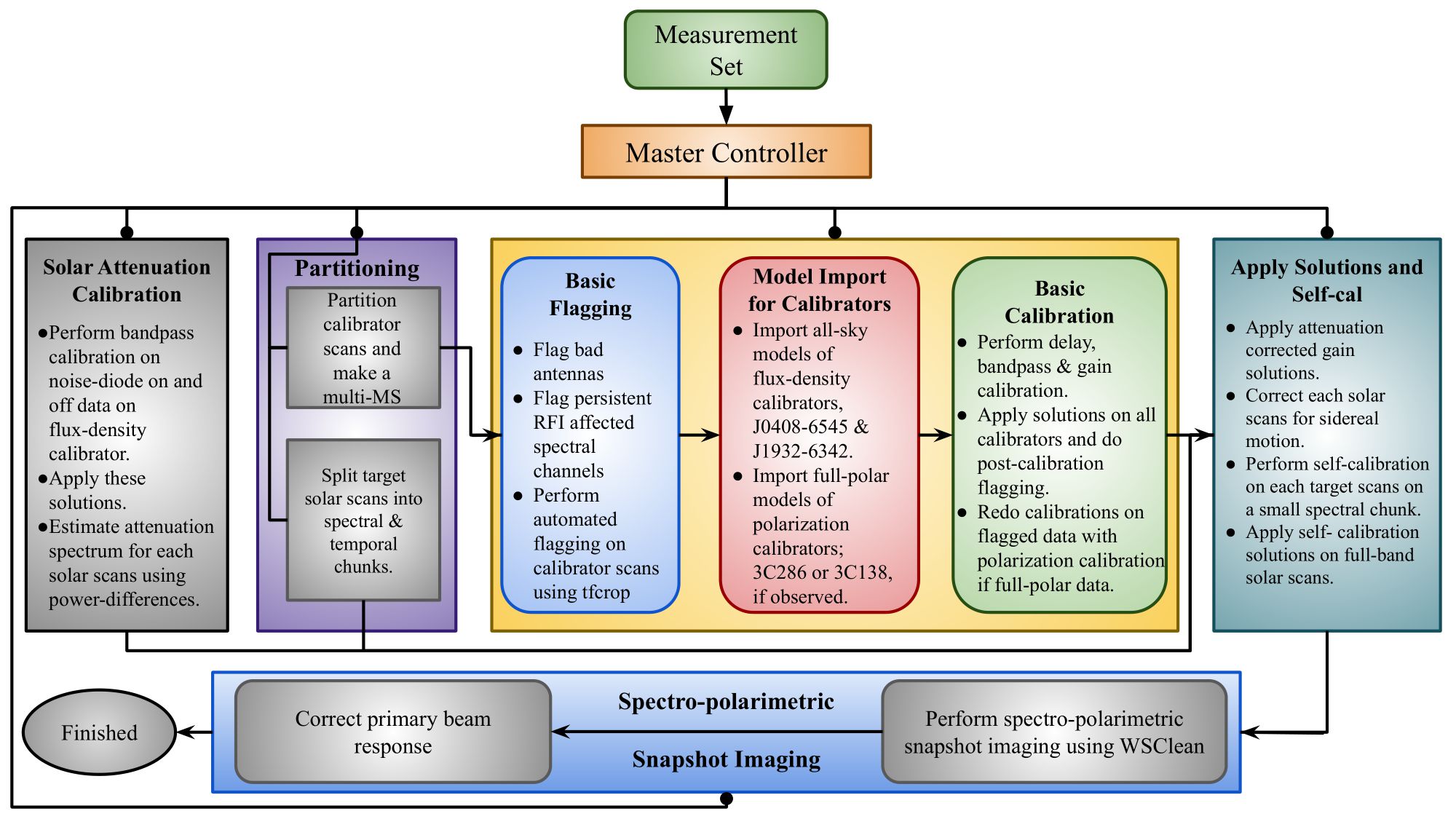}
    \caption{A flowchart of the pipeline for calibration and imaging of solar observations with MeerKAT. The master controller controls the workflow of the pipeline. Tasks inside individual rectangular blocks are executed in parallel and controlled by the master controller.}
    \label{fig:pipeline_schematic}
\end{figure*}

\section{Description of the Calibration and Imaging Pipeline}\label{sec:calibration}
Boresight solar observations with MeerKAT are non-standard and demand specialized calibration and imaging strategies. Existing tools such as \textsf{SolarKAT} \citep{Samboco2024}, designed to mitigate solar contamination in standard astronomical data, are not suitable for this purpose. To automate this process, we developed a dedicated pipeline inspired by \textsf{IDIA-processMeerKAT} \citep{IDIA_pipeline_2021}. This pipeline is fully automated, user-friendly, and deployable in both standard single-node workstations as well as high-performance cluster environments. It is distributed through PyPI \href{https://pypi.org/project/meersolar/}{https://pypi.org/project/meersolar/}. All images presented in this work are produced from observations automatically calibrated and imaged using this pipeline.

Its key features include:
\begin{enumerate}
    \item Support for both Full-Stokes (polarization) calibration and imaging of solar observations.
    \item Process-based parallelization using \textsf{Dask} \citep{dask-2015}, enabling cross-platform execution from clusters to single-node systems, unlike the MPI-based parallelism in \textsf{IDIA-processMeerKAT}.
    \item Efficient operation on memory-constrained machines, allowing large dataset processing where traditional tools may fail.
    \item Remote monitoring of pipeline progress.
\end{enumerate}

The pipeline uses \textsf{CASA} for calibration and \textsf{WSClean} for imaging. While examples of Stokes I imaging are presented here, full polar imaging will be presented in a forthcoming publication describing the imaging pipeline (Patra et al., in prep.). Figure \ref{fig:pipeline_schematic} illustrates the pipeline flowchart. A master controller manages modular blocks, with independent tasks (e.g., attenuation calibration, data partitioning) running in parallel, while sequential tasks (e.g., calibration, self-calibration, imaging) are executed in order. Internal parallelism within blocks, such as per-scan calibration steps and time-chunked self-calibration, maximizes computational efficiency.

\subsection{Data Partitioning, Flagging and Calibration}\label{subsec:basic_flagcal}
The Measurement Set (MS) is partitioned by scans and converted into multi-MS format for parallel processing using \textsf{Dask}. Flagging, calibration, and application of gain solutions are performed in parallel across scans. Persistent RFI and faulty antennas are flagged in all calibrator and solar scans. Automated RFI flagging using \textsf{flagdata} in \textsf{tfcrop} mode is applied to flux and phase calibrators but skipped for solar scans due to the intrinsic variability of solar emission. Bandpass calibrator models (\href{https://skaafrica.atlassian.net/wiki/spaces/ESDKB/pages/1481408634/Flux+and+bandpass+calibration\#Applying-a-full-sky-model-to-a-CASA-measurement-set}{MeerKAT flux density and bandpass calibrator models}) are used, and only scans without noise-diode firings are selected to derive delay, bandpass, and gain solutions using \textsf{gaincal} and \textsf{bandpass}, limited to baselines $>200\lambda$ to avoid contamination from large angular scale quiet Sun emission. Post-calibration flagging is applied to residuals using \textsf{rflag}, followed by a final calibration round. However, bright compact solar features may still contaminate the longer baselines used. To assess this, we shift the phase center of the calibrator scans to the Sun and generate a dirty image using baselines $>200\lambda$. If the resulting contamination level, quantified as $\Delta I / I$, exceeds 2\% (corresponding to a tolerable gain error of 1\%), we perform direction-dependent calibration to subtract the solar contribution and repeat the calibration iteration.

\begin{figure*}[!htbp]
    \centering
    \includegraphics[width=0.35\linewidth]{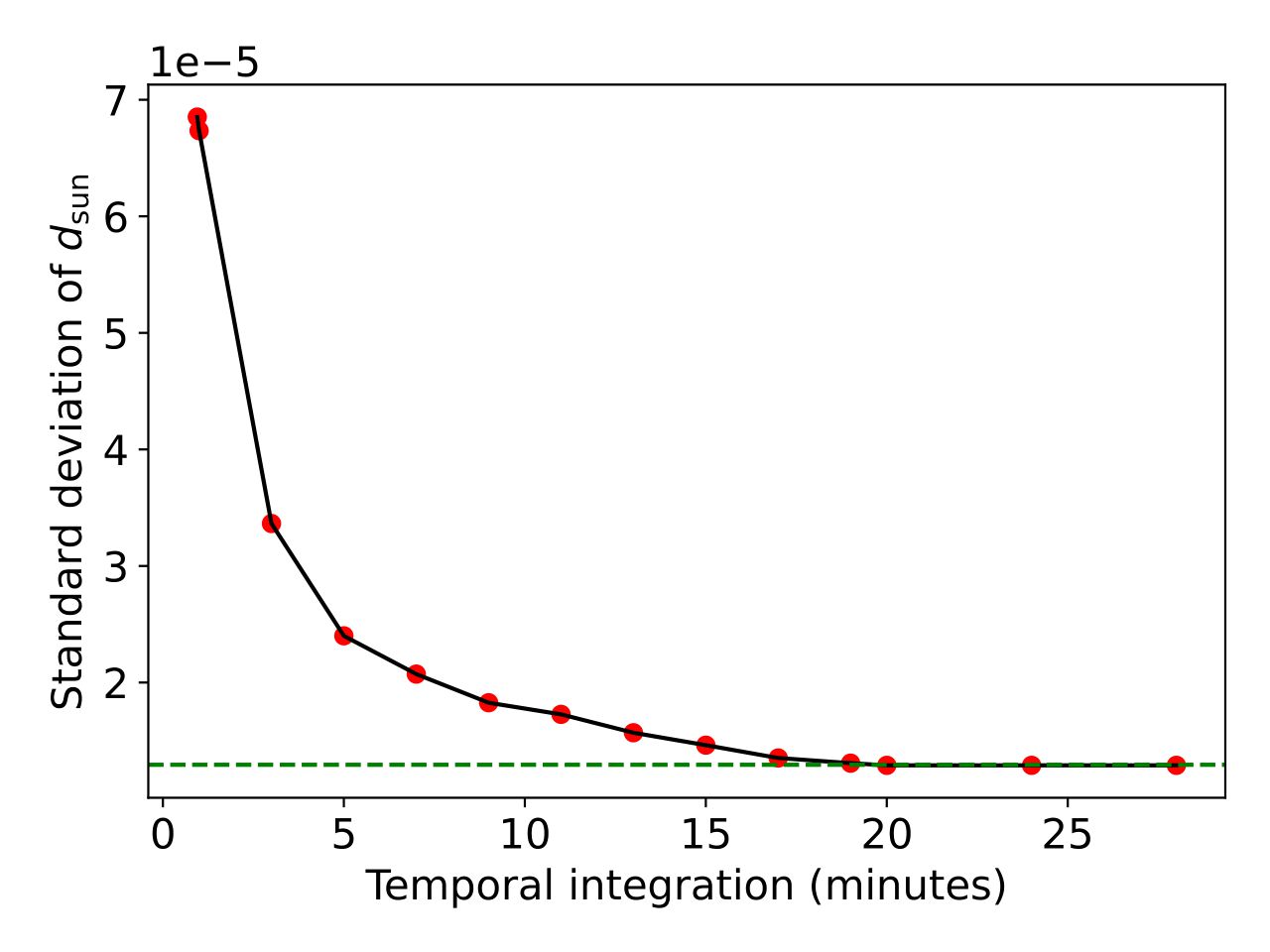}\includegraphics[trim={0.7cm 0.7cm 0cm 0cm},clip,width=0.33\linewidth]{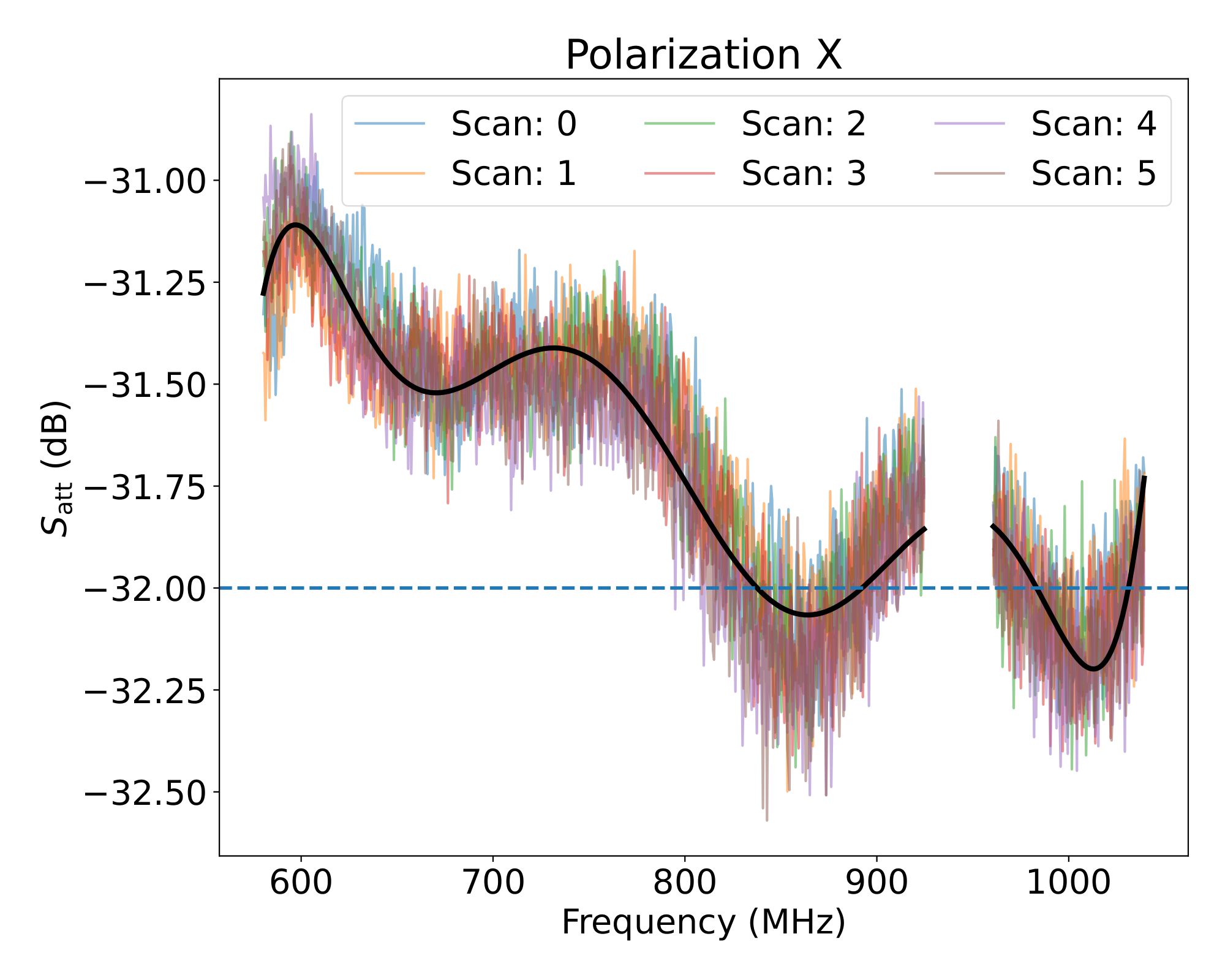}\includegraphics[trim={0.7cm 0.7cm 0cm 0cm},clip,width=0.33\linewidth]{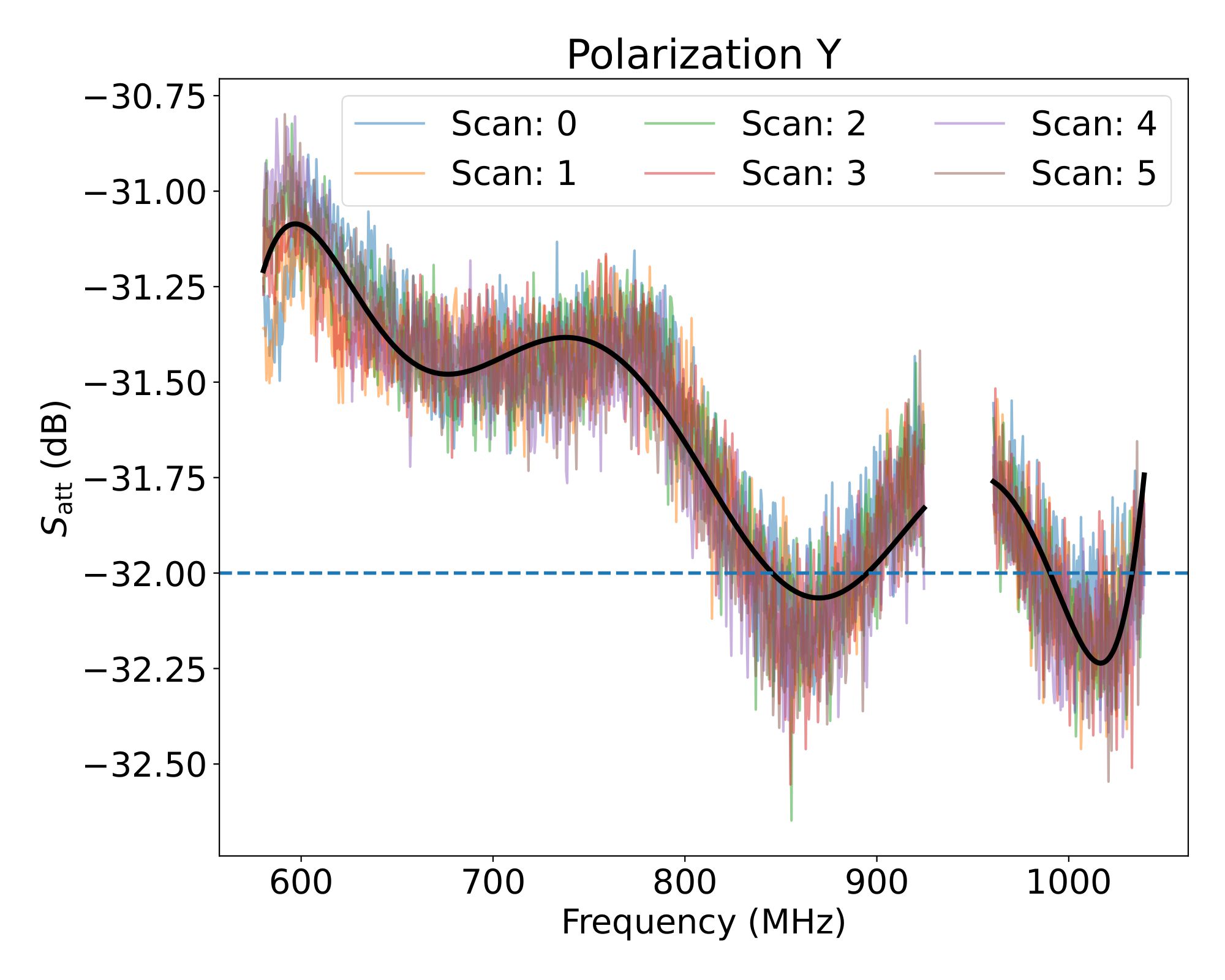}
    \caption{Left panel: Variation of band-averaged standard deviation of $d_\mathrm{sun}$ for different temporal integrations. It is evident, beyond 15 minutes of averaging, that the standard deviation $d_\mathrm{sun}$ does not drop further. Right panel: Spectrum of the estimated \( S_\mathrm{att} \) across different scans. The blue dashed line represents the fixed \( S_\mathrm{att} \) value applied during all scans. It is evident that the observed \( S_\mathrm{att} \) fluctuates within \( \pm 1 \) dB of the set value across scans. Solid black lines in the middle and right panels show the fitted response on the scan-averaged spectrum.}
    \label{fig:att_sun}
\end{figure*}

\subsection{Flux Density Calibration using Noise Diodes}\label{subsec:fluxcal_noisediode}
Absolute flux density calibration of solar observations is performed in two steps. First, the instrumental bandpass is calibrated without applying $S_\mathrm{att}$, using bandpass calibrator scans as described in Section \ref{subsec:basic_flagcal}. Next, the spectral response of $S_\mathrm{att}$ is calibrated using the noise diode by measuring the change in auto-correlation power between the diode-on and diode-off states in both calibrator ($d_\mathrm{cal}$) and solar ($d_\mathrm{sun}$) scans. For calibrator scans without $S_\mathrm{att}$, the diode induces a significant power increase, leading to an estimated $\sim$2\% variation in gain due to the departure of the system from linearity. In contrast, the effect is negligible for solar scans with $S_\mathrm{att}$ due to the fractionally much smaller power increase. To allow us to correct for this non-linearity, bandpass solutions are derived separately for the diode-on and diode-off states using the calibrator scans and applied accordingly.

While power variations due to the noise diode are easily detectable in calibrator scans without averaging, this is not the case for solar scans due to the suppression by $S_\mathrm{att}$. Individual $d_\mathrm{sun}$ estimates are noisy, but their SNR improves with integration time, %. The SNR of $d_\mathrm{sun}$ increases with integration time 
until they get limited by intrinsic solar variability. To determine the optimal averaging time, we evaluated the standard deviation of the $d_\mathrm{sun}$ spectrum as a function of integration time. As shown in the left panel of Figure \ref{fig:att_sun}, the standard deviation saturates beyond 15 minutes. Therefore, we adopt 15 minutes as the optimal integration interval for estimating $d_\mathrm{sun}$ from solar scans. The middle and right panels of Figure \ref{fig:att_sun} show the estimated $S_\mathrm{att}$ spectra for a UHF-band observation with 32 dB attenuation in both polarizations. The spectra exhibit intrinsic frequency dependence but no significant scan-to-scan variation beyond noise. To avoid using noisy per-channel estimates and increasing SNR, we fit a cubic spline to the scan-averaged spectra and use it to scale the bandpass solutions applied to the solar scans. The fitted spectra are shown by the solid black lines in the middle and right panels of Figure \ref{fig:att_sun}.

\subsection{Spectroscopic Snapshot Self-calibration}\label{subsec:selfcal}
Solar radio emission exhibits strong spectral and temporal variability, making the solar sky model inherently time-dependent. The time and frequency scale of the variation depends on solar activity and can range from a few seconds to several minutes and a few kHz to several MHz. To address this dynamic temporal and spectral variability scale, the spectral and temporal axes are adaptively divided into chunks such that deviations from the mean remain below certain thresholds (default is 10\% for frequency, 1\% for time). This ensures that variability is preserved while enabling computationally efficient self-calibration. A spectral chunk from the lower part of the band is selected to maximize surface brightness sensitivity and improve modeling on shorter baselines. The self-calibration procedure follows the convergence criteria in \citet{Kansabanik2022_paircarsI,Kansabanik_paircars_2}, starting with phase-only calibration and advancing to joint amplitude-phase calibration upon convergence. CLEAN thresholding is progressively reduced, and the process is stopped when no further improvement in image dynamic range is observed. Convergence is defined as a relative change in dynamic range below a user-defined threshold $\epsilon$ over three iterations, with a maximum iteration cap to prevent oscillatory behavior for small $\epsilon$.

\subsection{Spectroscopic Snapshot Imaging}\label{subsec:imaging}
At GHz frequencies, solar radio emission exhibits structure across a broad range of angular scales—from arcseconds to the full solar disc—often with significant complexity. The imaging pipeline supports user-defined baseline selection and weighting strategies, and by default adopts Briggs weighting \citep{briggs1995} with a robust parameter of 0.0 to achieve a balance between resolution and sensitivity. Multiscale deconvolution is employed with frequency-dependent multiscale parameters. These choices, detailed in Appendix \ref{app:multiscale_params}, are made to avoid deconvolution artifacts.

\begin{figure*}[!htbp]
\centering
    \includegraphics[trim={0cm 0cm 0.2cm 0cm},clip,width=0.89\linewidth]{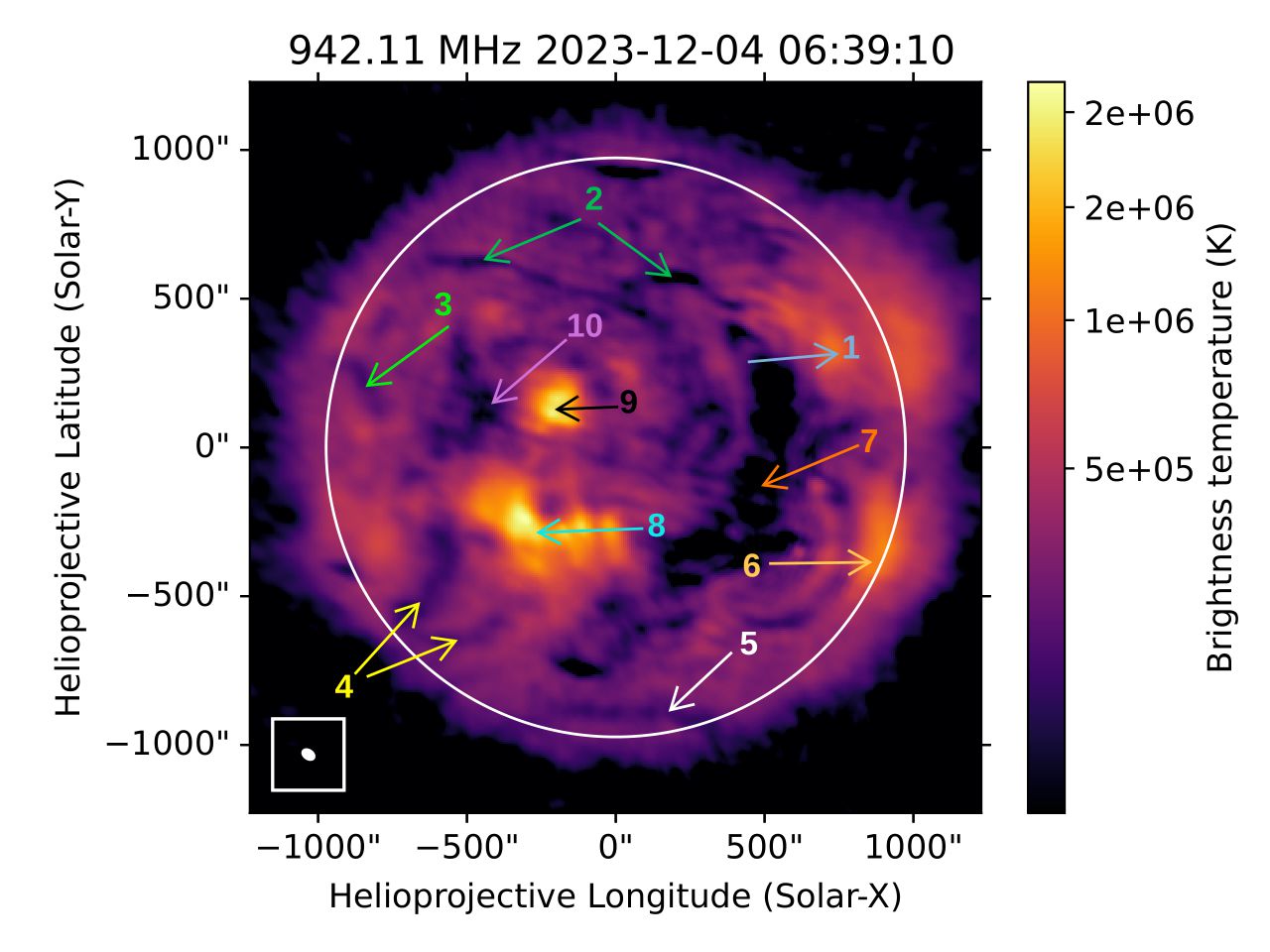}\\
    \includegraphics[trim={0cm 0cm 3.6cm 0.7cm},clip,scale=0.72]{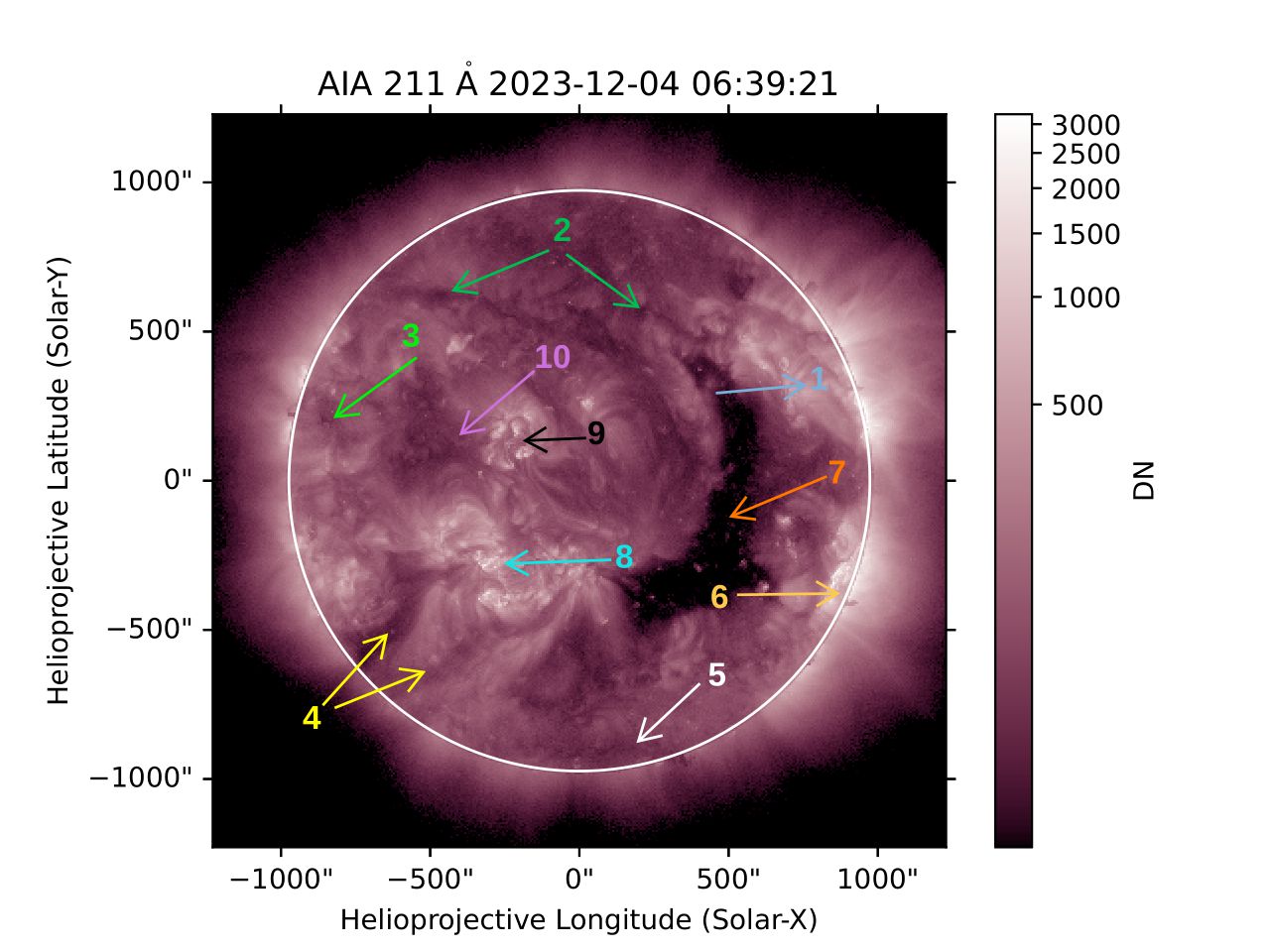}\includegraphics[trim={2.5cm 0cm 1cm 0.7cm},clip,scale=0.72]{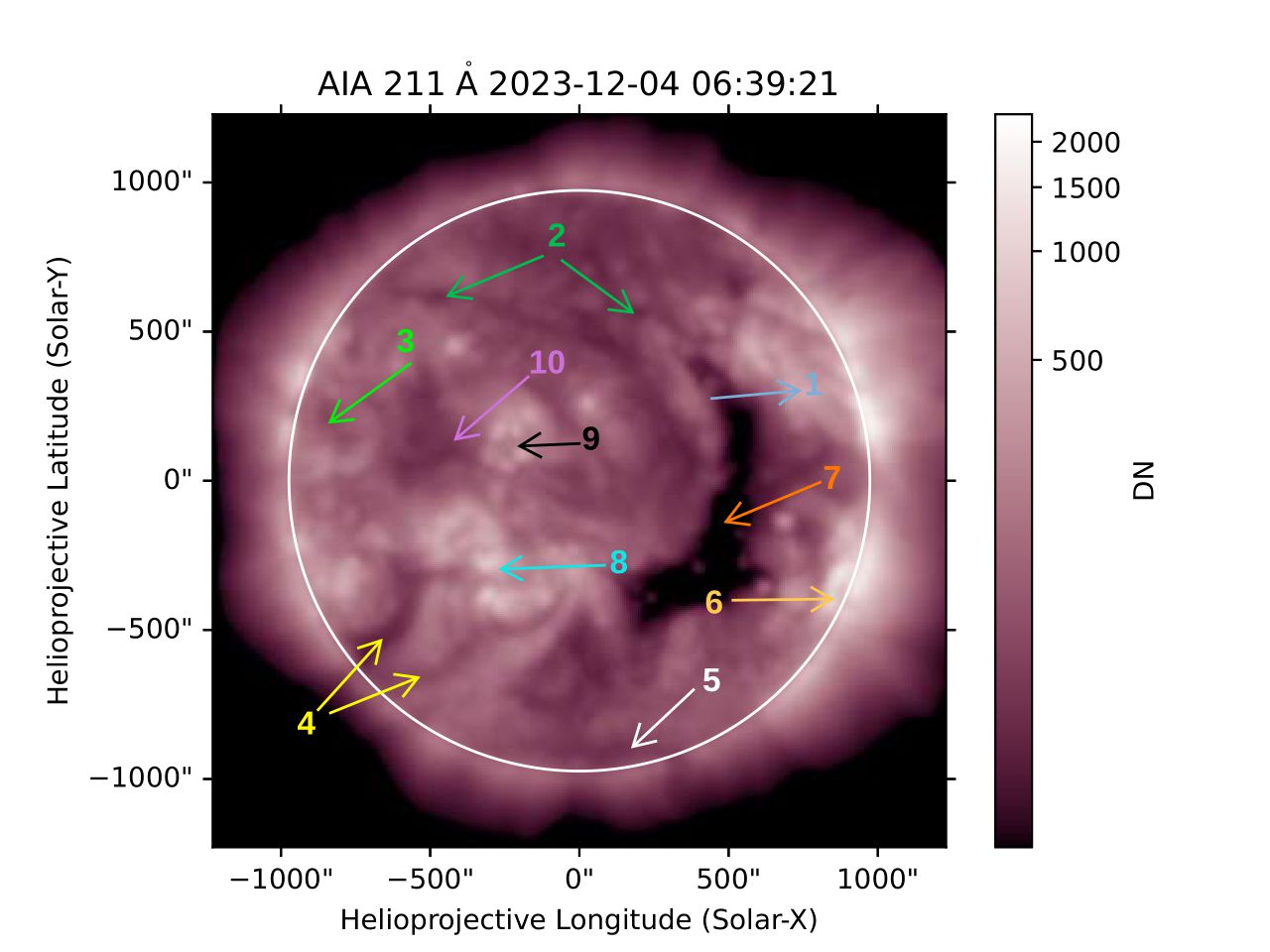}
    \caption{Morphological comparison between MeerKAT UHF band image with 211\AA\ EUV image from SDO/AIA. The top panel shows the radio image from MeerKAT (averaged data over 50 MHz spectral and 15 minutes of temporal chunk) showing emissions at more than $5\sigma$ significance. The bottom left is the AIA image at its original spatial resolution ($1.2''$), and the right image is the AIA image convolved with a Gaussian beam of size similar to the PSF of the radio image ($18''$). Several features have been marked by numbered arrows in all three images for visual guidance.}
    \label{fig:morphological_compaison}
\end{figure*}

\subsection{Primary Beam Correction}\label{subsec:pb_cor}
As the Sun is an extended source, its observed emission must be corrected for the direction-dependent primary beam response. We apply image-based primary beam correction using the array-averaged MeerKAT beam model from holography measurements \citep{deVilliers_2022,deVilliers2023}. The beam is described by the Jones matrix $P(l, m, \nu)$ in direction cosines ($l,\ m)$) from the boresight of the telescope. At MeerKAT, $H$ and $V$ correspond to $Y$ and $X$ polarizations (\href{https://katdal.readthedocs.io/en/latest/signs.html\#polarisation}{MeerKAT polarization convention}) as per IAU convention \citep{IAU_1973} used in common softwares like with \textsf{CASA} and \textsf{WSClean}. $H$ and $V$ polarizations are appropriately labeled in the IAU convention in the measurement set using \href{https://katdal.readthedocs.io/}{katdal} software package. Hence, appropriate changes are also made in $P(l, m, \nu)$ to be consistent with the IAU convention.

To apply correction, the beam is first mapped from $(l, m)$ to equatorial coordinates of the image, then rotated by the parallactic angle $\chi$. The sky-frame beam matrix is:
\begin{equation}
    P_\mathrm{sky}=P(l,m)\ R(\chi)
\end{equation}
where the parallactic rotation matrix is:
\begin{equation}
    R(\chi)=\begin{pmatrix}
\mathrm{cos}\chi & -\mathrm{sin}\chi \\
\mathrm{sin}\chi & \mathrm{cos}\chi 
\end{pmatrix}
\end{equation}
The frequency-averaged Stokes I beam is computed as:
\begin{equation}
    P_\mathrm{I}=\frac{\sum_{\nu=\nu_0}^{\nu_1}\left[\frac{1}{2}\sum_{i,j=0}^{i,j=1}|P_\mathrm{sky}[i,j]|^2\right]}{(\nu_1-\nu_0)}
\end{equation}
where $\nu_0$ and $\nu_1$ represent the start and end frequencies of the image. The Stokes I image corrected for the primary beam is obtained by dividing by $P_\mathrm{I}$. We note that while UHF and L-band have overlapping frequencies, the primary beam should be derived from the appropriate band, as they use different feeds, resulting in distinct beam responses even at the same frequency.

\begin{figure*}[!htbp]
    \includegraphics[trim={0cm 1cm 0.8cm 0cm},clip,width=0.53\linewidth]{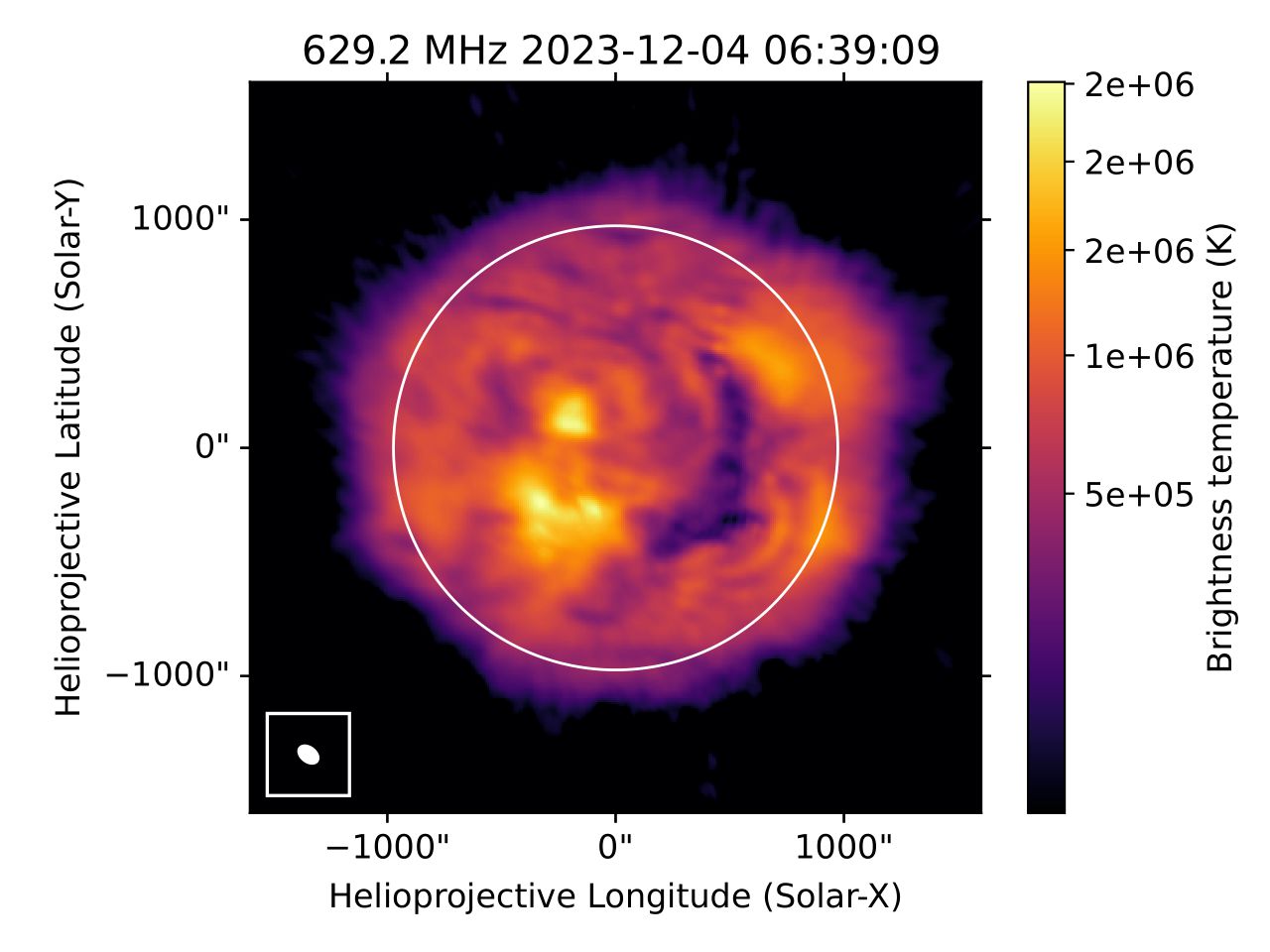}\includegraphics[trim={3cm 1cm 0cm 0cm},clip,width=0.46\linewidth]{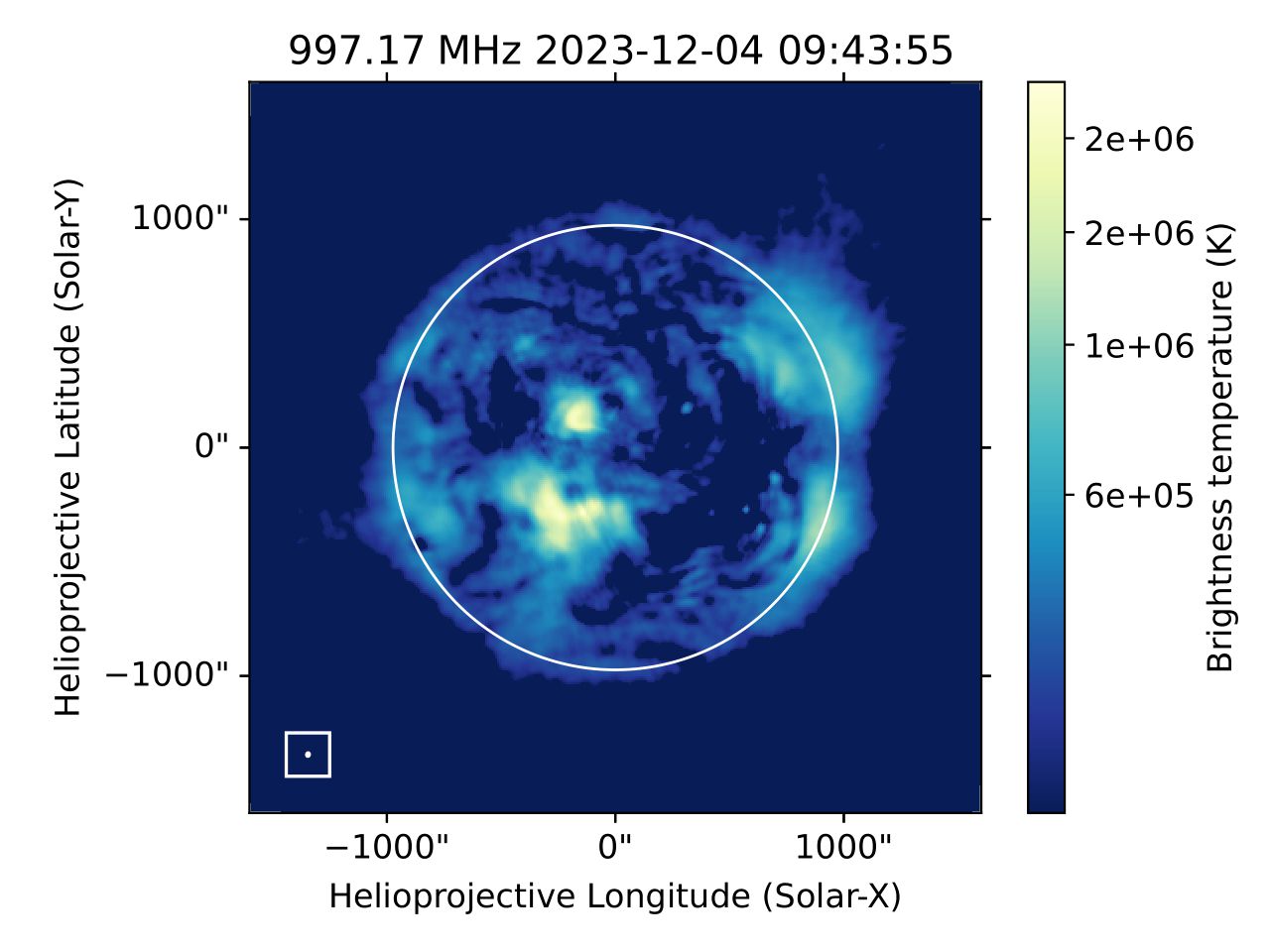}
    \includegraphics[trim={0cm 1cm 0.8cm 0.3cm},clip,width=0.53\linewidth]{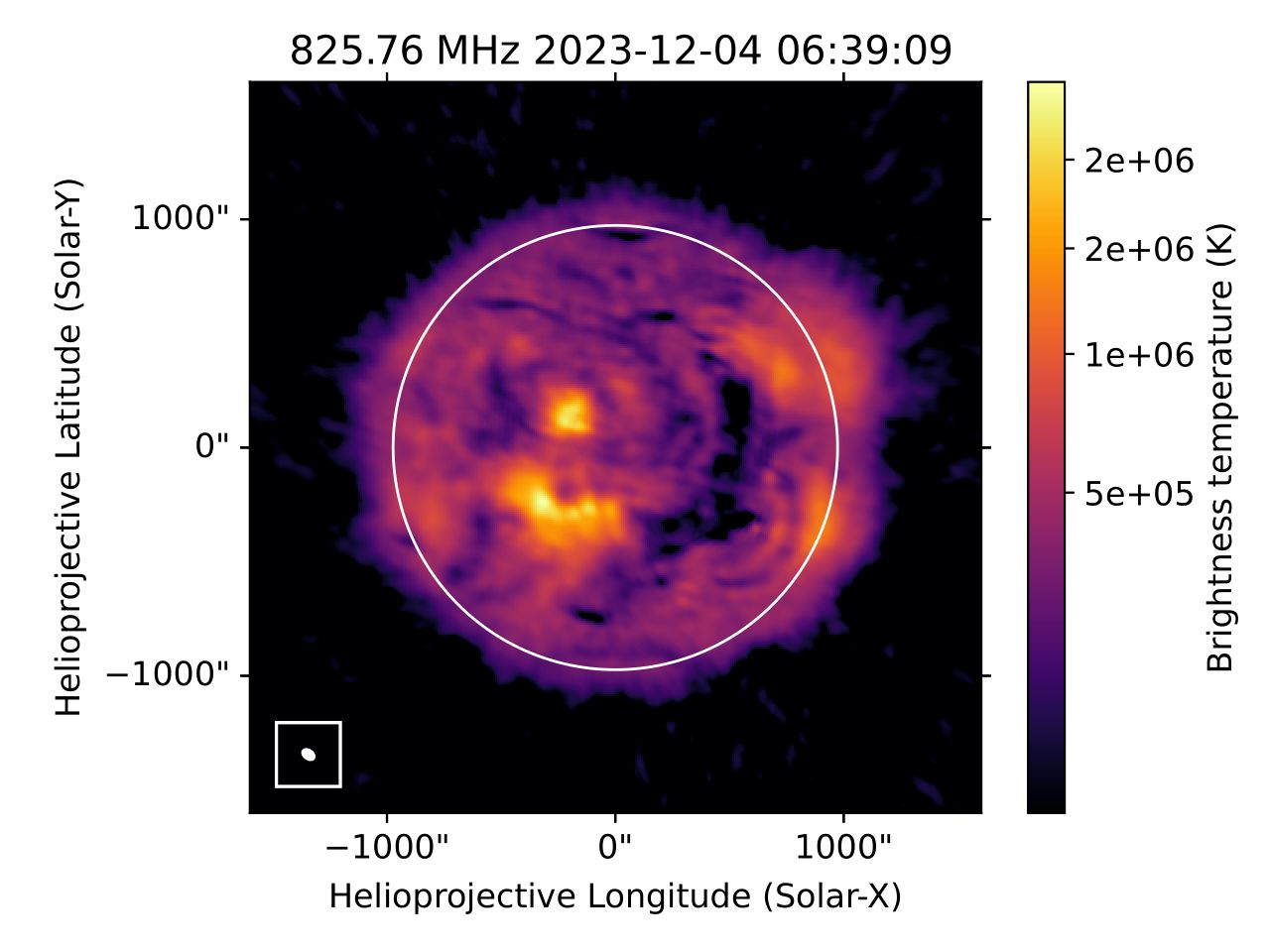}\includegraphics[trim={3cm 1cm 0cm 0cm},clip,width=0.46\linewidth]{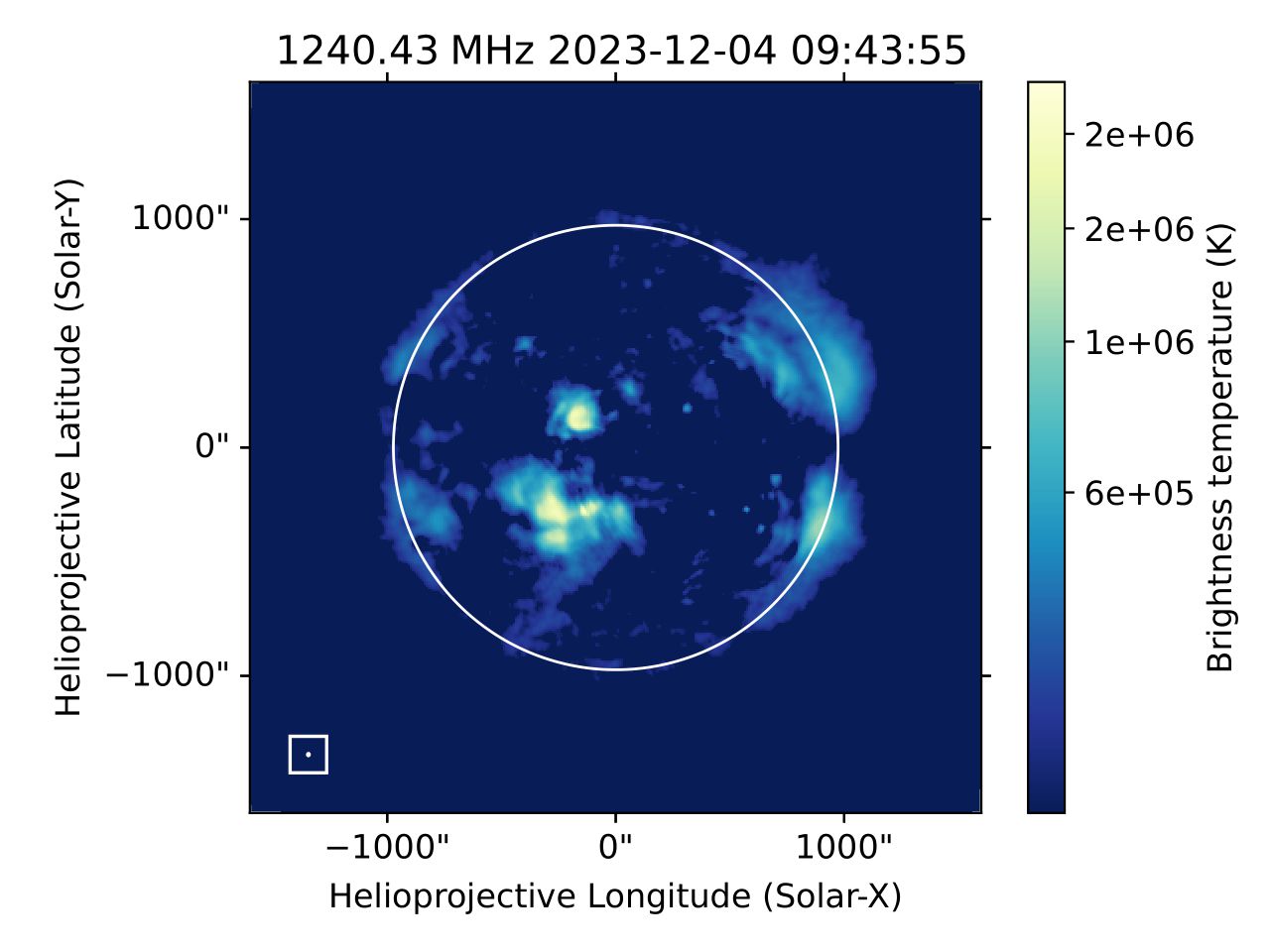}
    \includegraphics[trim={0cm 0cm 0.8cm 0.3cm},clip,width=0.53\linewidth]{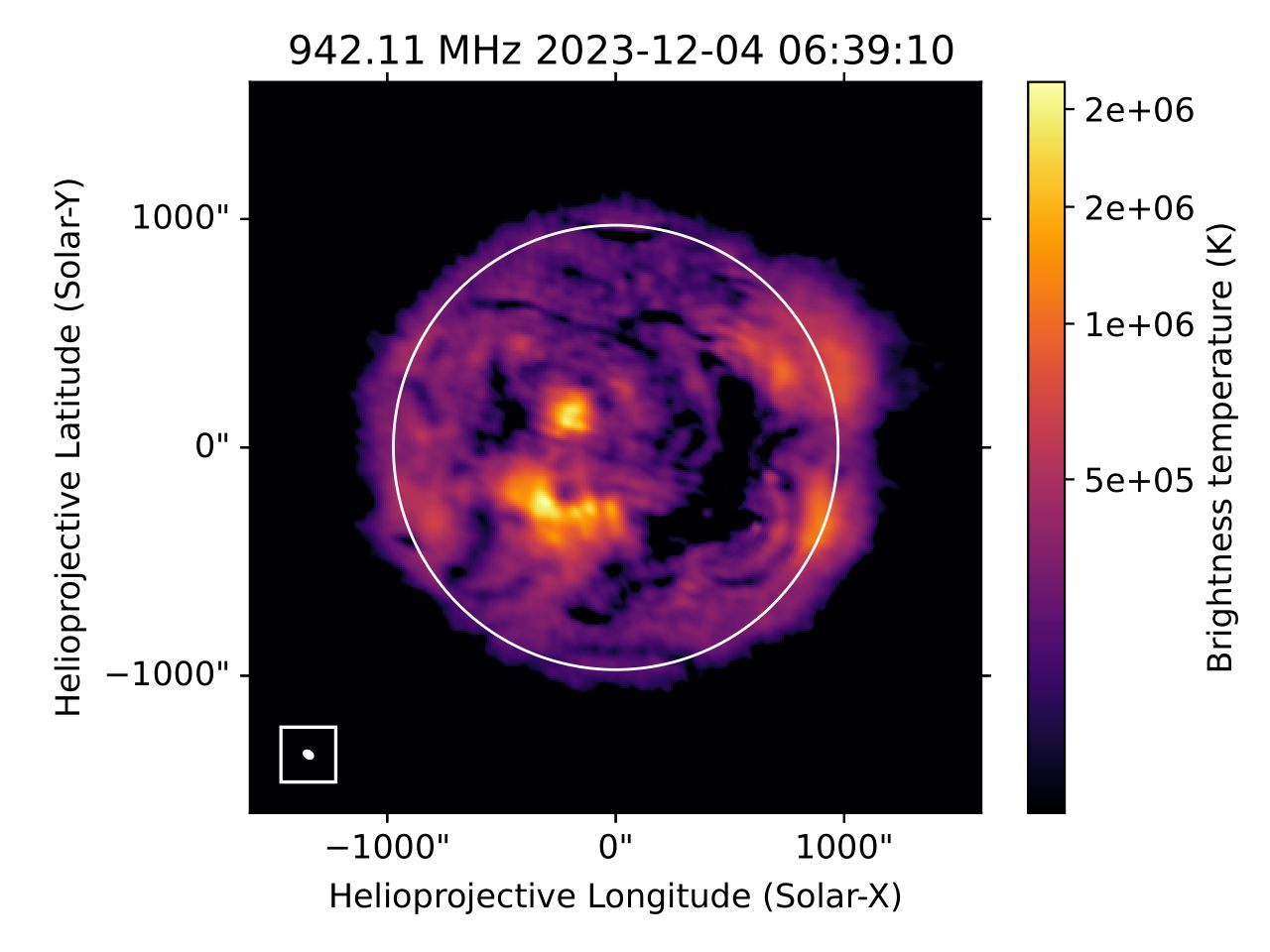}
    \includegraphics[trim={3cm 0cm 0cm 0cm},clip,width=0.46\linewidth]{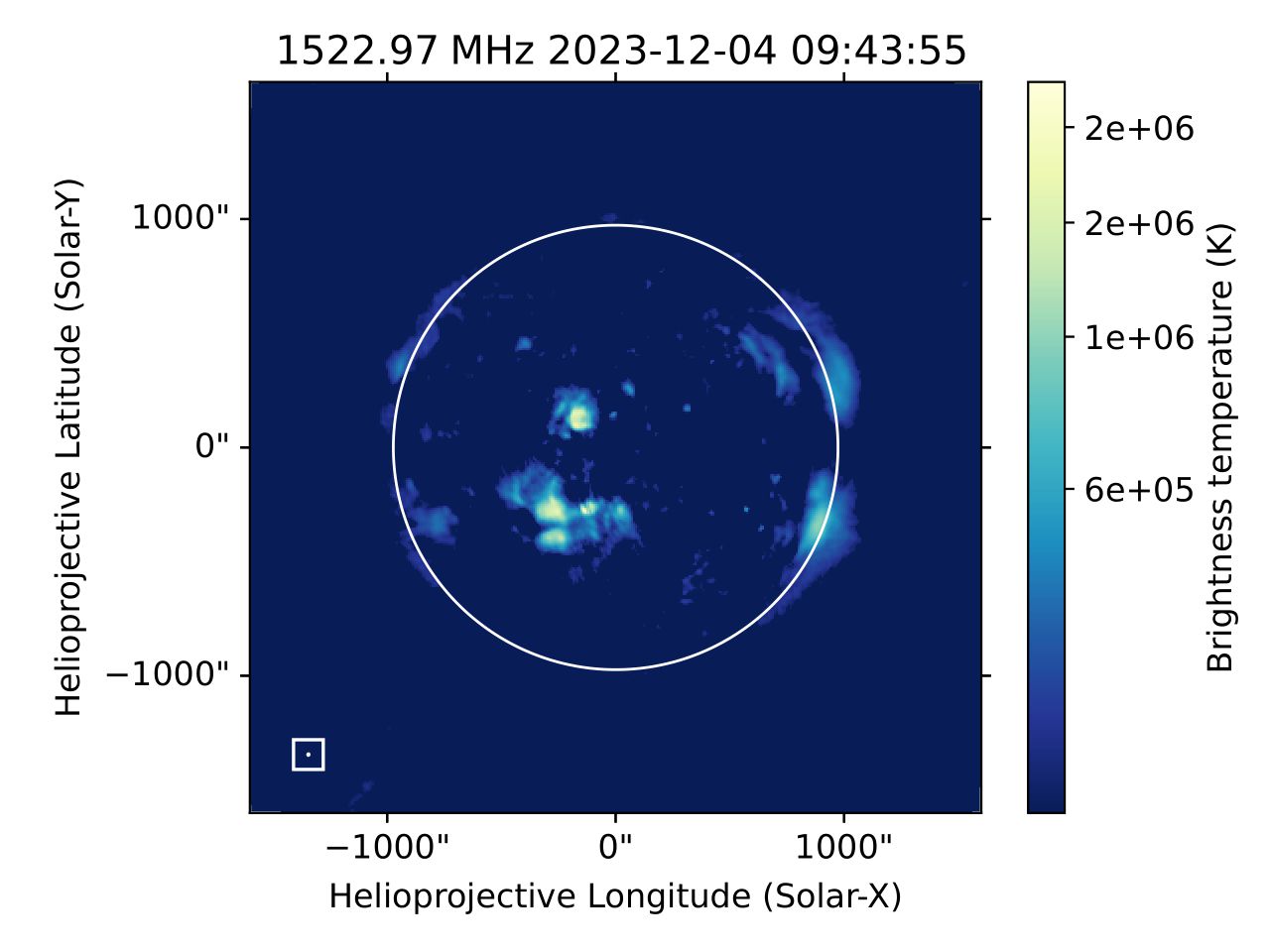}
    \caption{Spectroscopic radio images from MeerKAT observations on 4 December 2023 are shown (50 MHz and 15 minutes integrations). The left and right panels display images from the UHF and L-band observations, respectively, above $5\sigma$ detection significance at three representative frequencies. The white circles indicate the optical solar disc. Synthesized beams are shown at the bottom left of each image, which varies between $\sim25''$ at the lowest frequency at UHF-band to $\sim5''$ at the highest frequency in the L-band.}
    \label{fig:spectroscopic_imaging}
\end{figure*}

\section{Observation Details}\label{sec:observation}
MeerKAT, with its field of view of roughly $\sim1^{\circ}$ \citep{deVilliers2023}, enables full-disk and lower-corona solar observations in a single pointing at the solar center, eliminating the need for mosaicking. In the standard solar observing mode, visibilities are recorded with 4096 spectral channels and a temporal resolution of 2~s. The excellent spectroscopic and snapshot {\it uv}-coverage supports high-quality imaging at these native resolutions, while the final choice of spectral and temporal averaging can be tailored by the user based on scientific requirements and available computing resources.
 
We have used data taken as part of engineering tests (project ID: EXT-20221114-PK-01) and the SARAO (South African Radio Astronomy Observatory) Science Verification (SSV) observations (project ID: SSV-20240609-DK-01). Engineering tests are performed in both UHF- and L-band, while SSV observations were taken only in the UHF-band.  Results from observations performed on 04 December 2023 and 10 June 2024 are presented in this paper. All spectroscopic snapshot images presented in this paper are generated using a 50 MHz bandwidth and 15 minutes of temporal averaging for the ease of data analysis.

\section{System Verification}\label{sec:verify_system}
To evaluate the accuracy of calibration and image reconstruction in the presence of attenuators, we compare radio images with co-temporal extreme ultraviolet (EUV) observations. Figure \ref{fig:morphological_compaison} shows this comparison using EUV images from Atmospheric Imaging Assembly \citep[AIA,][]{Lemen2012} onboard Solar Dynamics Observatory \citep[SDO;][]{pesnell2012}, closest in time to the radio image. The MeerKAT image centered at 942.11 MHz is produced using 50 MHz and 15 minutes of data. Key features are highlighted with colored arrows, which are identical across all panels. The large coronal hole (region 7) appears with similar morphology in both bands. An additional, smaller coronal hole is also marked (region 10). On-disk active regions (regions 8 and 9) and eastern limb active regions (regions 1 and 6) are identifiable. Additional filamentary structures are indicated by regions 2, 3, 4, and 5.

Despite the high surface brightness sensitivity of MeerKAT, its shortest baseline is not short enough to detect structures comparable in size to the solar disc at the upper end of the L-band, the minimum baseline length of approximately $160\lambda$ corresponds to the largest angular scale of $\sim22'$, resulting in missing flux from extended coronal structures \citep{Kansabanik_2024_meerkat} at frequencies higher than $\sim900$ MHz. However, below 900 MHz, the shortest baseline ($\sim$29 m) of MeerKAT is capable of capturing emission at an angular scale $\sim40'$. Hence, at the UHF band, we do not expect significant missing flux, as already demonstrated by \cite{Kansabanik_2024_meerkat}. Figure \ref{fig:spectroscopic_imaging} presents spectroscopic images from UHF and L-band observations taken on 2024 December 04. The UHF-band images (left panels) capture extended diffuse emission, whereas the L-band images (right panels) show a loss of large-scale structure toward the higher end of the frequency range. All of these images have spectral and temporal integrations of 50 MHz and 15 minutes, respectively. The observed brightness temperatures reach values on the order of a million Kelvin, consistent with typical coronal temperatures. Notably, the near-overlapping frequency images from the UHF and L-band (top right and bottom left) exhibit similar morphology and peak brightness temperatures, indicating consistency in both flux density calibration and primary beam correction across the two bands.

\section{Glimpses of Early Science Results}\label{sec:early_science_results}
At metre wavelengths, high-quality spectro-polarimetric snapshot imaging with SKA-low precursors like the MWA and pathfinders such as LOFAR and uGMRT has already demonstrated the ability to observe a wide range of solar phenomena \citep{Oberoi2023}— from the quiet Sun \citep[e.g.,][]{Vocks2018,Sharma_2020,zhang2022}, coronal holes \citep{McCauley2019,Rahaman2019} and weak transient events \citep{Sharma2018,Mondal2020b,Mondal2021b}, to faint emissions from CME plasma \citep{Mondal2020a,Kansabanik_cme1,Kansabanik2024_cme2}, high-resolution imaging of solar noise storms \citep{Mondal_2024_gmrt,Mondal2025}, intense active emissions from high energy particles \citep[e.g.,][]{Mohan2019a,Mohan2021a,Mohan2021b}, CME shocks \citep[e.g.,][]{Bhunia2023,Zhang2024,Kumari2025}. Recent high-fidelity polarimetric studies \citep[e.g.,][]{McCauley2019,Rahaman2019,Morosan2022,Dey2025} also started providing new insights on the radio emission from the solar corona, $\sim 1.1-2.5\ R_\odot$, \citep{Gary1989}.

MeerKAT solar observations enable unprecedented spectroscopic snapshot imaging of the Sun at centimeter wavelengths. The high image fidelity, as demonstrated through comparisons with EUV images from AIA, will advance the study of the solar corona and eruptive events and open up discovery potential. This section provides glimpses of a range of new science objectives that can potentially be achieved using MeerKAT solar observations already available. Detailed analyses and discussion of individual science targets with higher spectro-temporal resolution are deferred to forthcoming publications.

\subsection{A Complementary Diagnostic of Multi-thermal Solar Atmospheric Plasma}\label{subsec:radio_dem}
EUV spectral lines \citep{Khan2022} and soft X-ray observations are widely used to probe the thermal structure of the solar atmosphere. Slit-based spectrographs like EUV Imaging Spectrometer onboard Hinode \citep[Hinode/EIS;][]{Culhane2007} and Coronal Diagnostic Spectrometer onboard Solar and Heliospheric Observatory \citep[SOHO/CDS;][]{Del2001} offer good spectral resolution and broad temperature coverage (log $T \sim 4.9\ –\ 6.5$), primarily focused on coronal plasma. The Interface Region Imaging Spectrograph \citep[IRIS;][]{DePontieu2014} provides better spectral and spatial resolution for chromosphere and transition region (TR) plasma. Although these instruments can probe all the layers of the solar atmosphere, their limited field of view (FoV) restricts them to studying local dynamics.

In contrast, full-disk EUV imagers such as SDO/AIA, Solar Ultraviolet Imager onboard Geostationary Operational Environmental Satellite \citep[GOES/SUVI;][]{goesuvi2022} and mid- and near-UV imagers like the Solar Ultraviolet Imaging Telescope onboard Aditya-L1 \citep[SUIT/Aditya-L1;][]{Tripathi2025}, provide high-cadence observations in multiple broad UV channels. While EUV-imagers nominally span log $T \sim 3.7\ –\ 8.0$, their temperature sensitivity for chromospheric and TR plasma is limited, and DEMs can be reliably estimated primarily in the log $T \sim 5.0\ –\ 8.0$ range using optically thin lines \citep{hannah2012,Cheung_2015}. 

Radio observations directly measure the free–free continuum emission from all of the plasma along the line of sight (LoS), providing sensitivity to the total emission measure across a broad temperature range. Hence, high-fidelity spectroscopic and snapshot imaging observation using MeerKAT offers a complementary tool to probe the full-disk TR and coronal plasma dynamics, filling the gap between slit-based spectrographs and EUV imagers. 
 
\begin{figure*}[!htbp]
\centering
\includegraphics[trim={0.5cm 0.5cm 0.4cm 0.3cm},clip,width=0.5\linewidth]{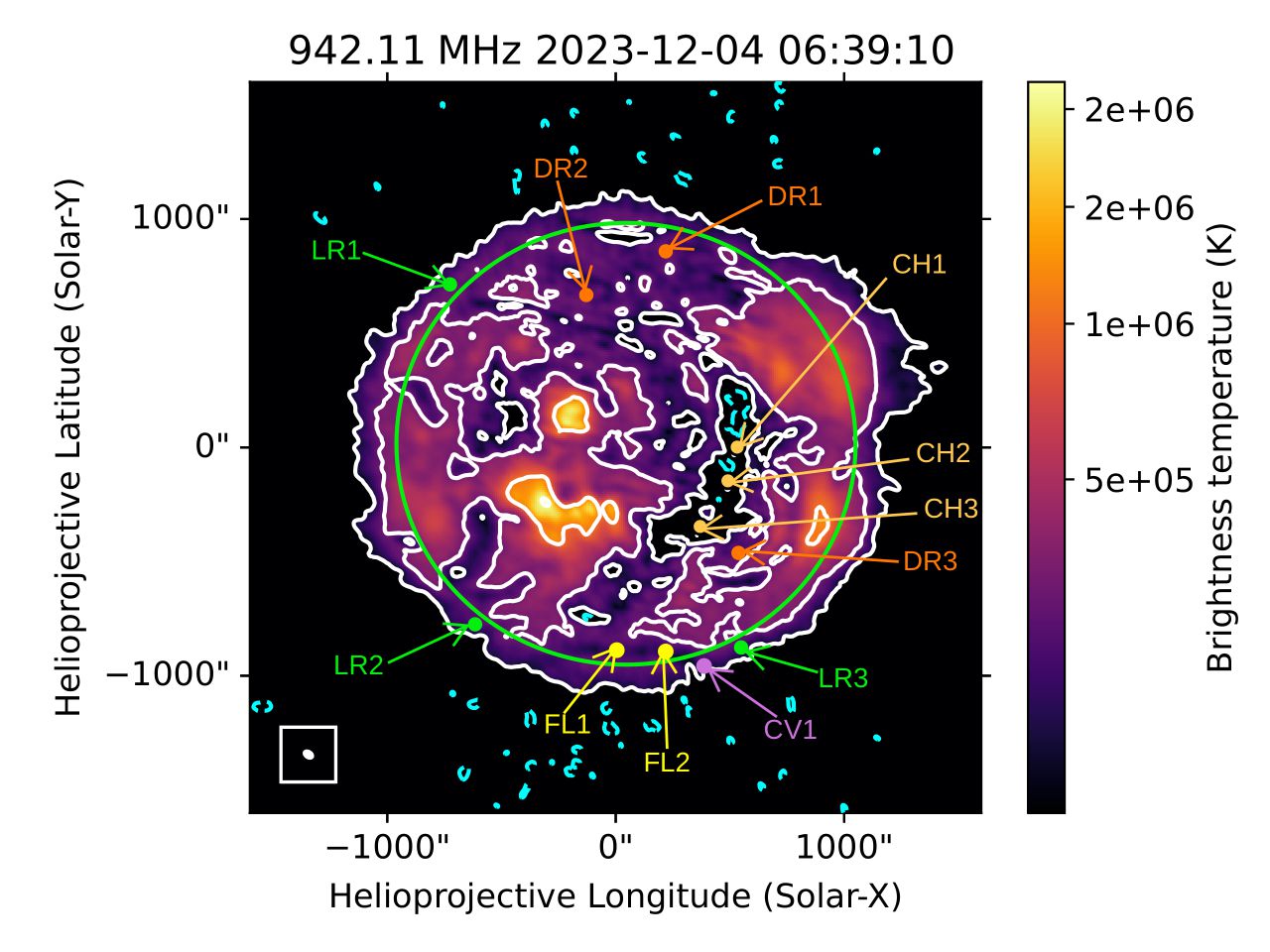}\includegraphics[trim={0.2cm 0.5cm 0cm 0cm},clip,width=0.5\linewidth]{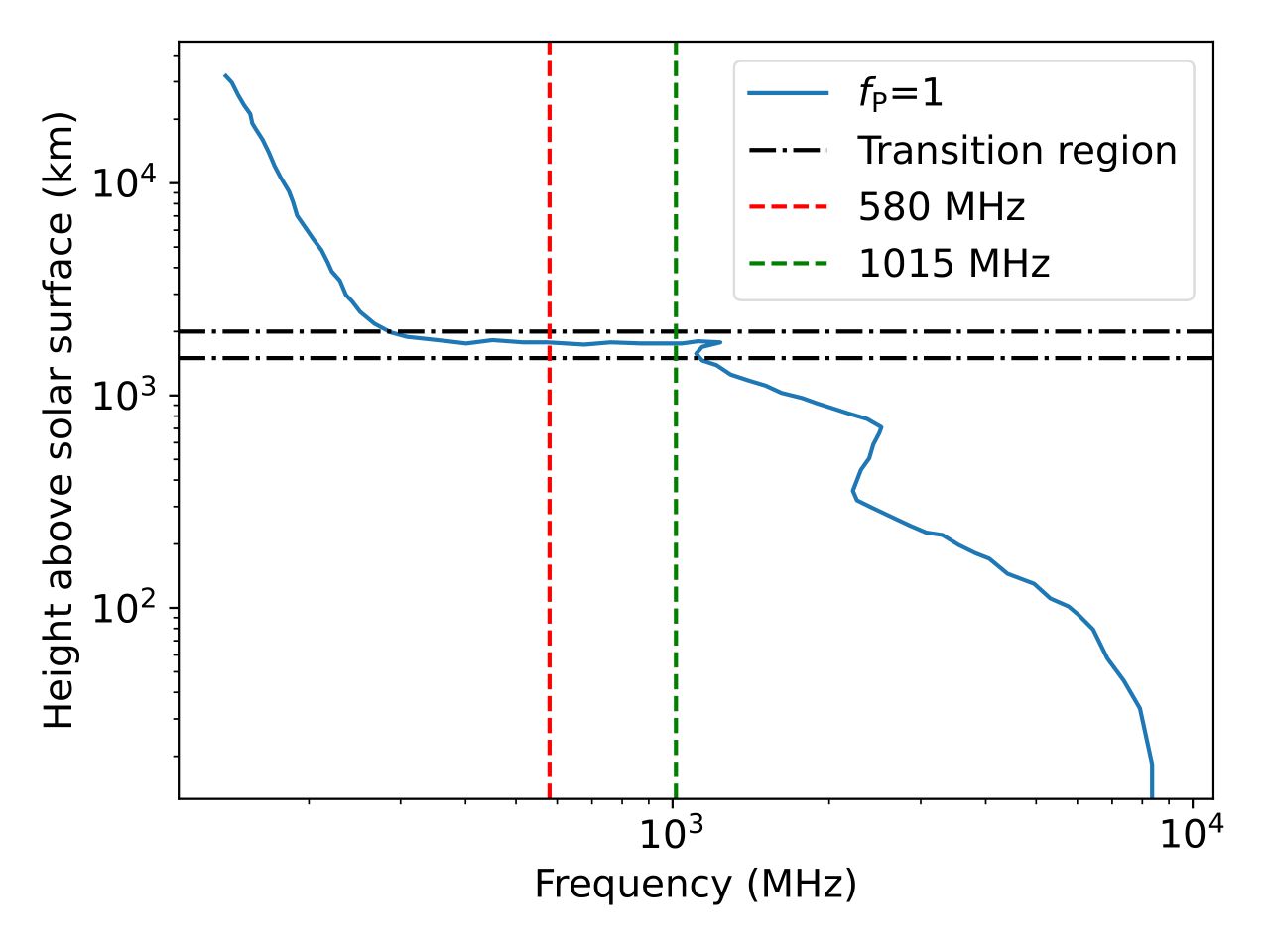}
\caption{The left panel shows an image from MeerKAT at 942 MHz on 2023 December 04. The PSF is shown at the bottom left of the image. The contour levels are at $-2.5, -5, 3, 15, 30, 60$ and $90\%$ of the peak brightness temperature. White contours represent positive values, and cyan contours represent negative values. A green circle marks the optical disk of the Sun. Different PSF-sized regions for which spectra are shown in Figure \ref{fig:spectra} are marked. The right panel shows plasma frequency ($f_\mathrm{P}$) as a function of height above the solar surface. Emission at MeerKAT UHF-band (580-1015 MHz) probes the atmospheric heights above the chromosphere.}
\label{fig:reg_plasma_freq}
\end{figure*}

\subsubsection{Study of of Quiescent Sun Plasma above the Chromosphere}\label{subsec:non_ar}
Radio waves cannot escape from regions where the local plasma frequency exceeds the emission frequency. Since the local plasma frequency is inversely related to local electron density and electron density varies with altitude above the solar surface, each frequency probes down to a specific atmospheric depth. Resolving the height structure of the solar atmosphere with spectroscopic radio imaging requires high spectral resolution and a broad frequency coverage across the characteristic plasma frequencies ($f_\mathrm{P}$) of the different layers. The highest frequency of MeerKAT UHF band in general lies at the bottom of the TR, as shown in the right panel of Figure \ref{fig:reg_plasma_freq}, determined using the electron density distribution of the solar atmosphere obtained from \cite{Aschwanden2005}. Hence, using MeerKAT UHF-band, which can capture diffuse quiescent emission well, probes emission above the chromosphere --- from the TR and corona. 

High-frequency ($>1$ GHz; JVLA, EOVSA, NoRH) and low-frequency ($\sim$150–432 MHz; NRH) instruments have long been used to study solar atmospheric layers \citep{Alissandrakis2020}, but their sparse \textit{uv} coverage and limited surface brightness sensitivity hinder imaging of faint, extended quiescent structures. Consequently, spatially resolved studies have focused on bright sources like active regions and flares \citep[e.g.,][]{Vourlidas_Bastian_1996,Bastian1998,Gary2018}, while quiet Sun analyses rely mostly on disk-integrated or spectrally averaged observations \citep[e.g.,][]{Zhang2001,Landi2008}. 

\begin{figure*}[!htbp]
\includegraphics[trim={0cm 0cm 0cm 0cm},clip,width=0.33\linewidth]{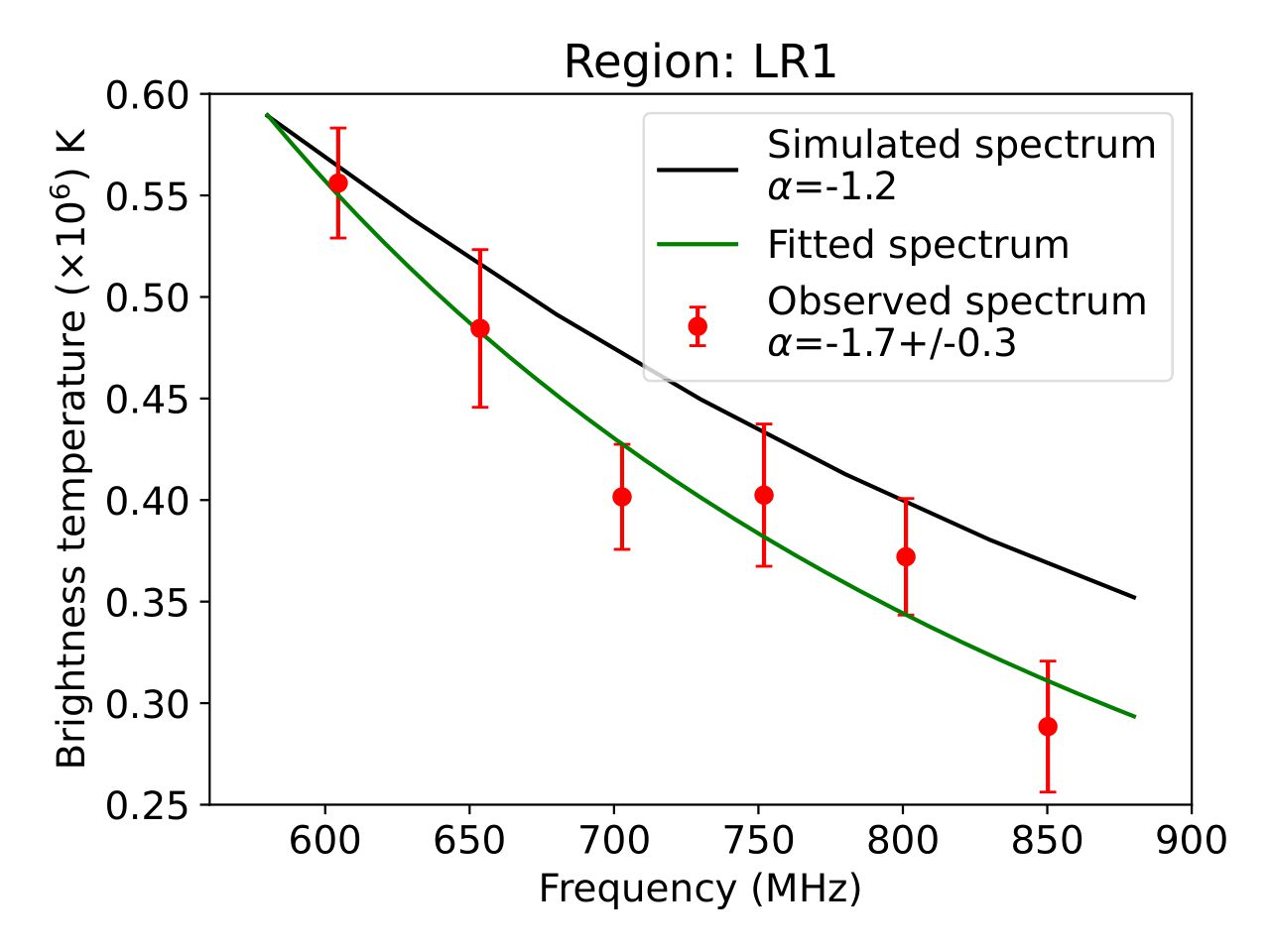}\includegraphics[trim={0cm 0cm 0cm 0cm},clip,width=0.33\linewidth]{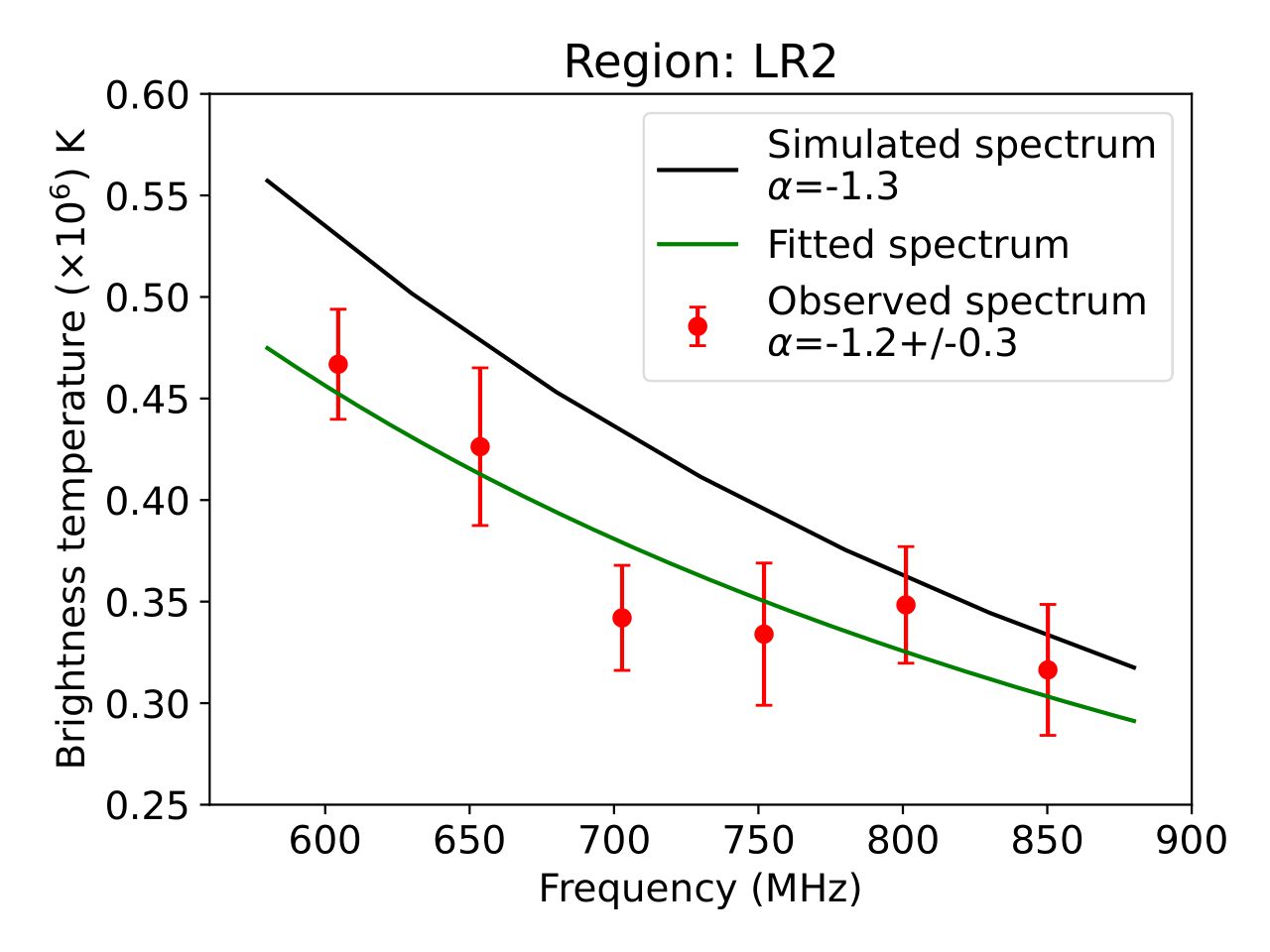}\includegraphics[trim={0cm 0cm 0cm 0cm},clip,width=0.33\linewidth]{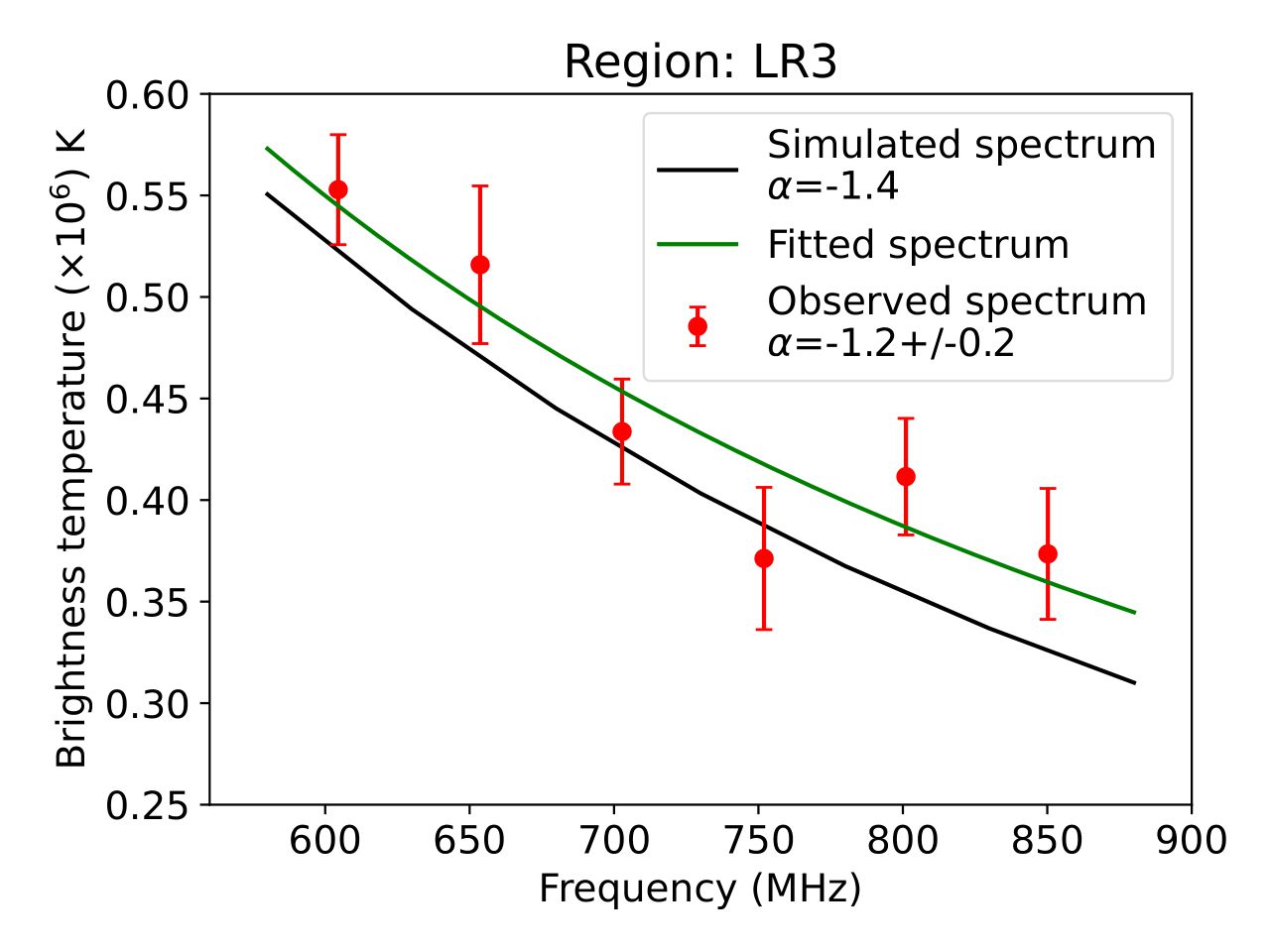}\\
\includegraphics[trim={0cm 0cm 0cm 0cm},clip,width=0.33\linewidth]{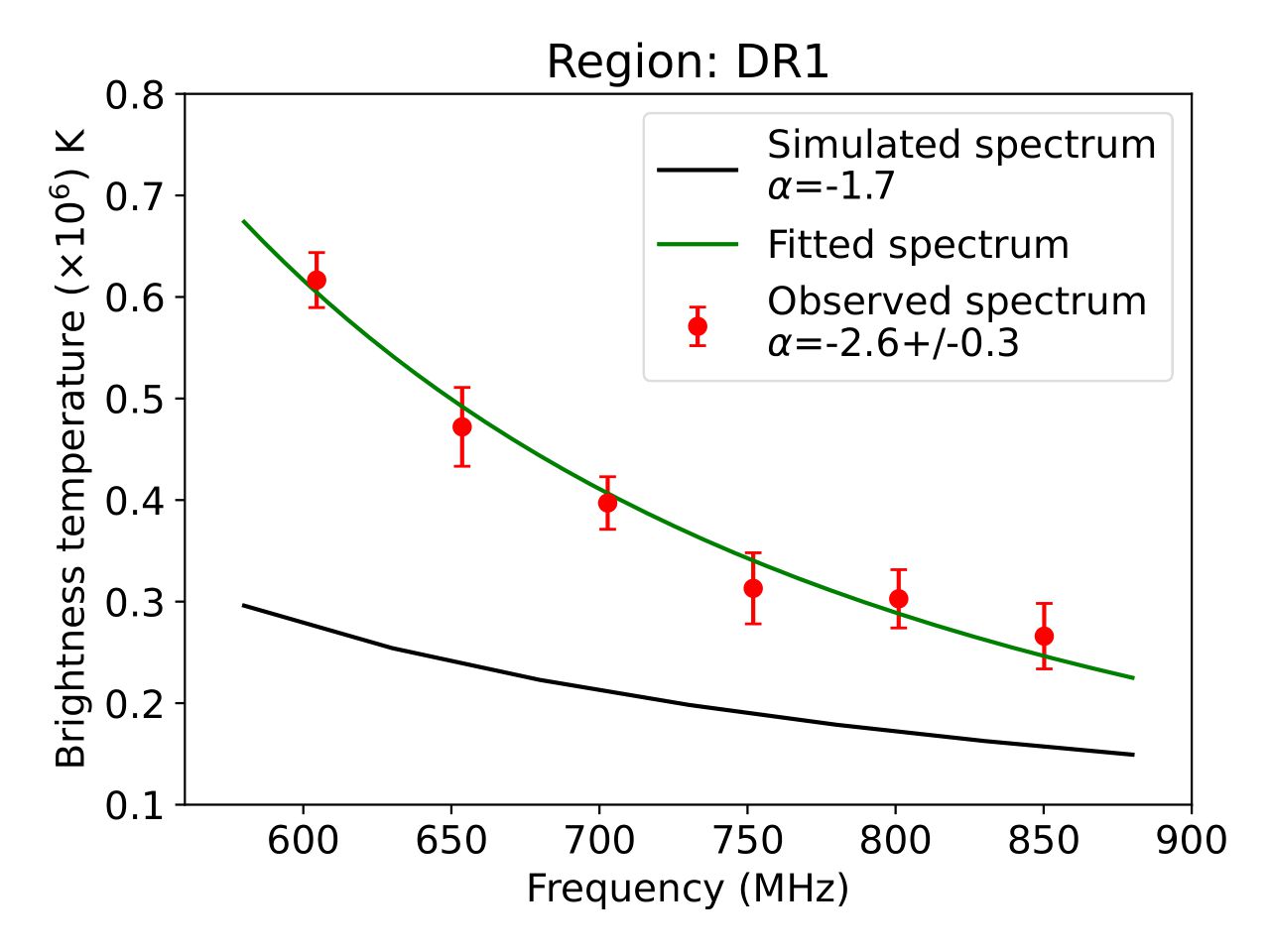}\includegraphics[trim={0cm 0cm 0cm 0cm},clip,width=0.33\linewidth]{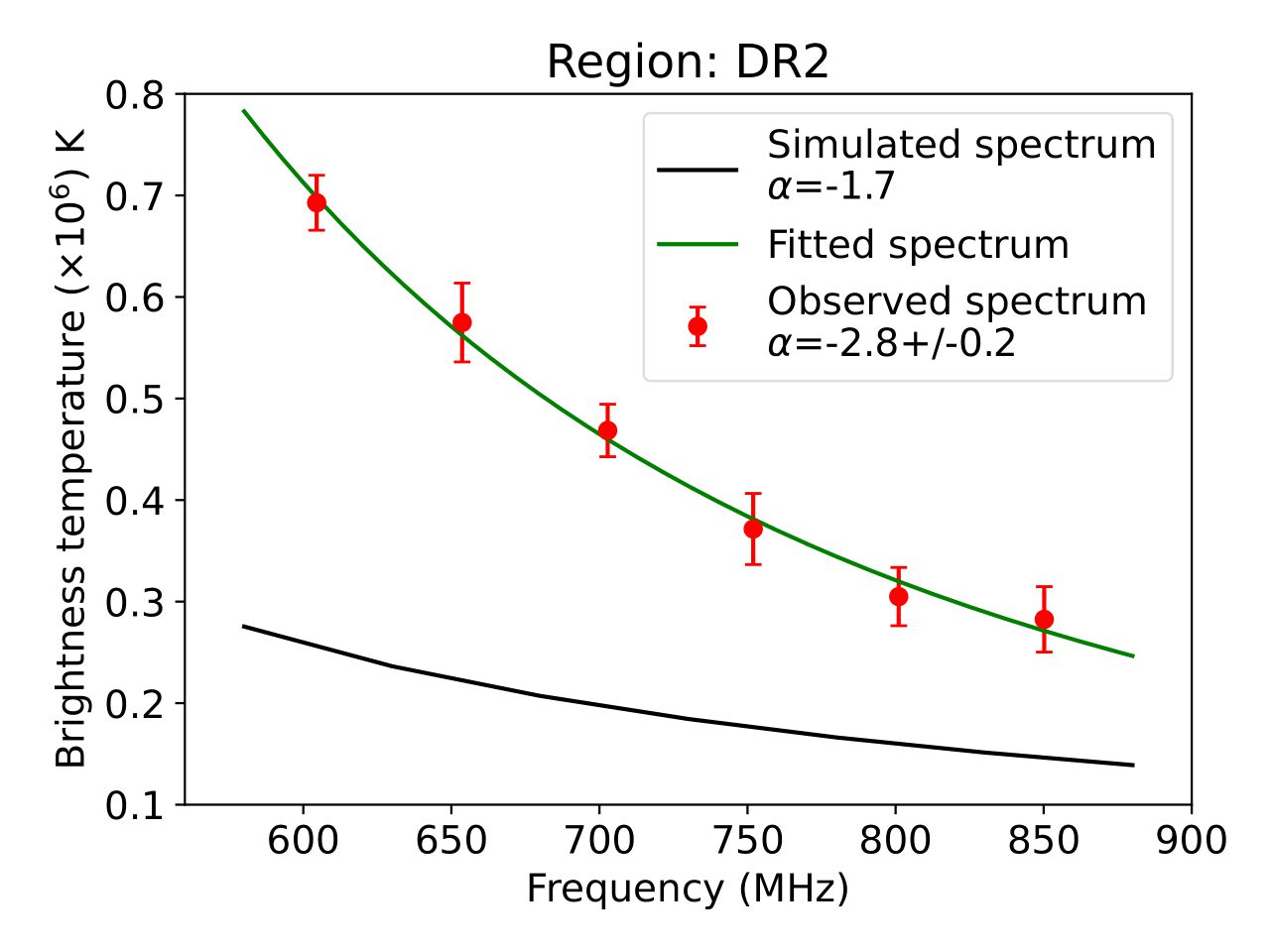}\includegraphics[trim={0cm 0cm 0cm 0cm},clip,width=0.33\linewidth]{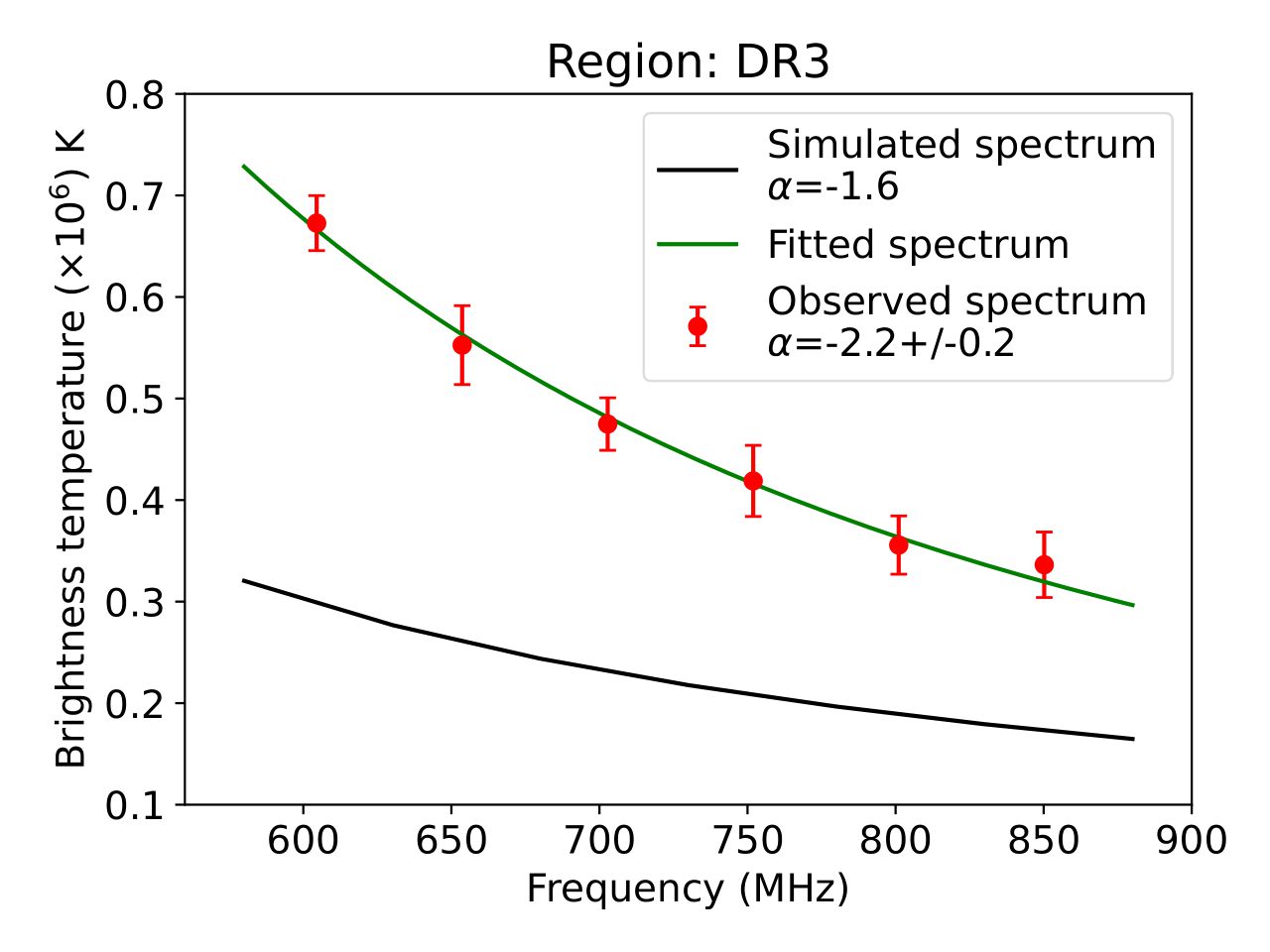}\\
\includegraphics[trim={0cm 0cm 0cm 0cm},clip,width=0.33\linewidth]{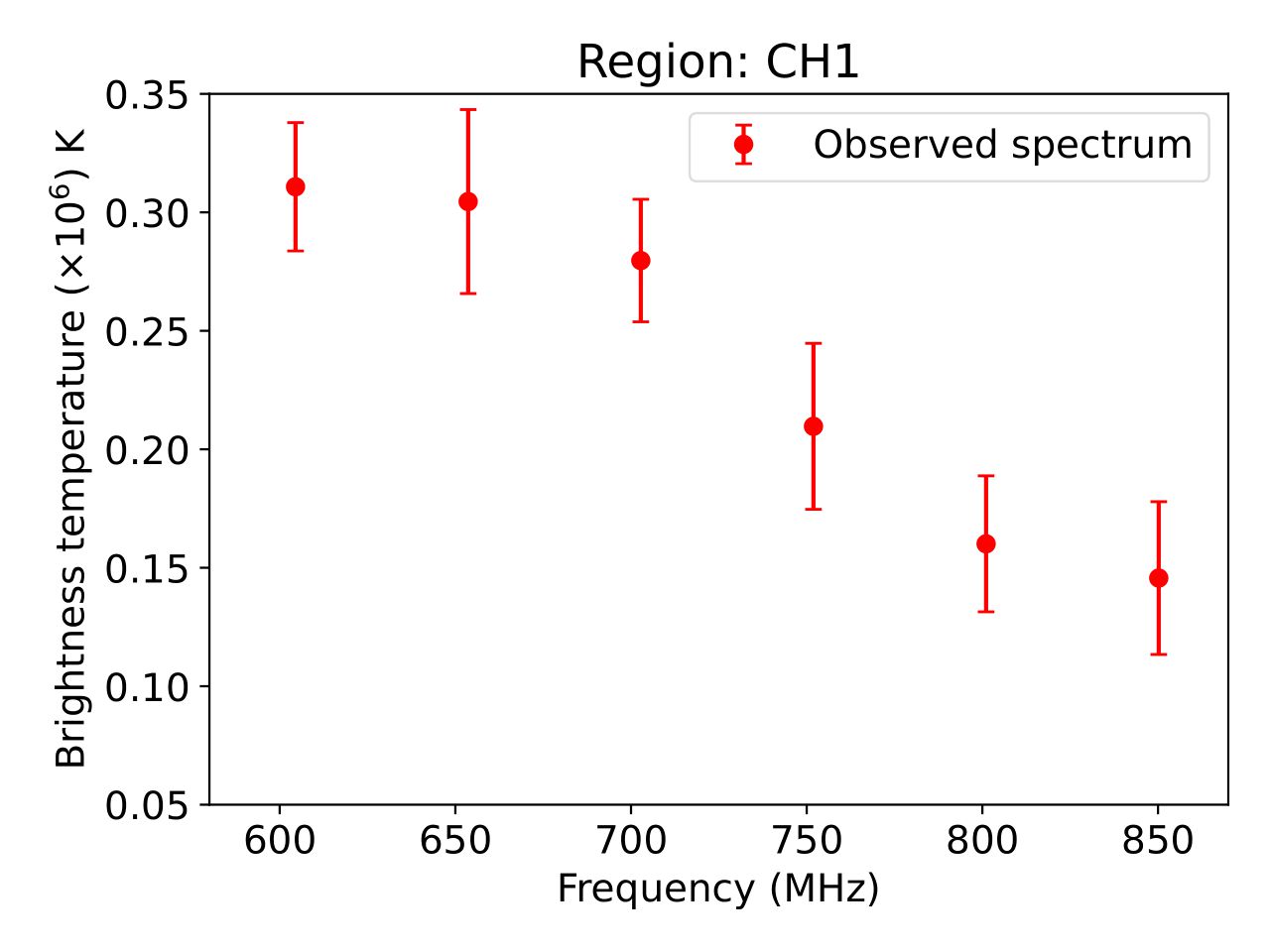}\includegraphics[trim={0cm 0cm 0cm 0cm},clip,width=0.33\linewidth]{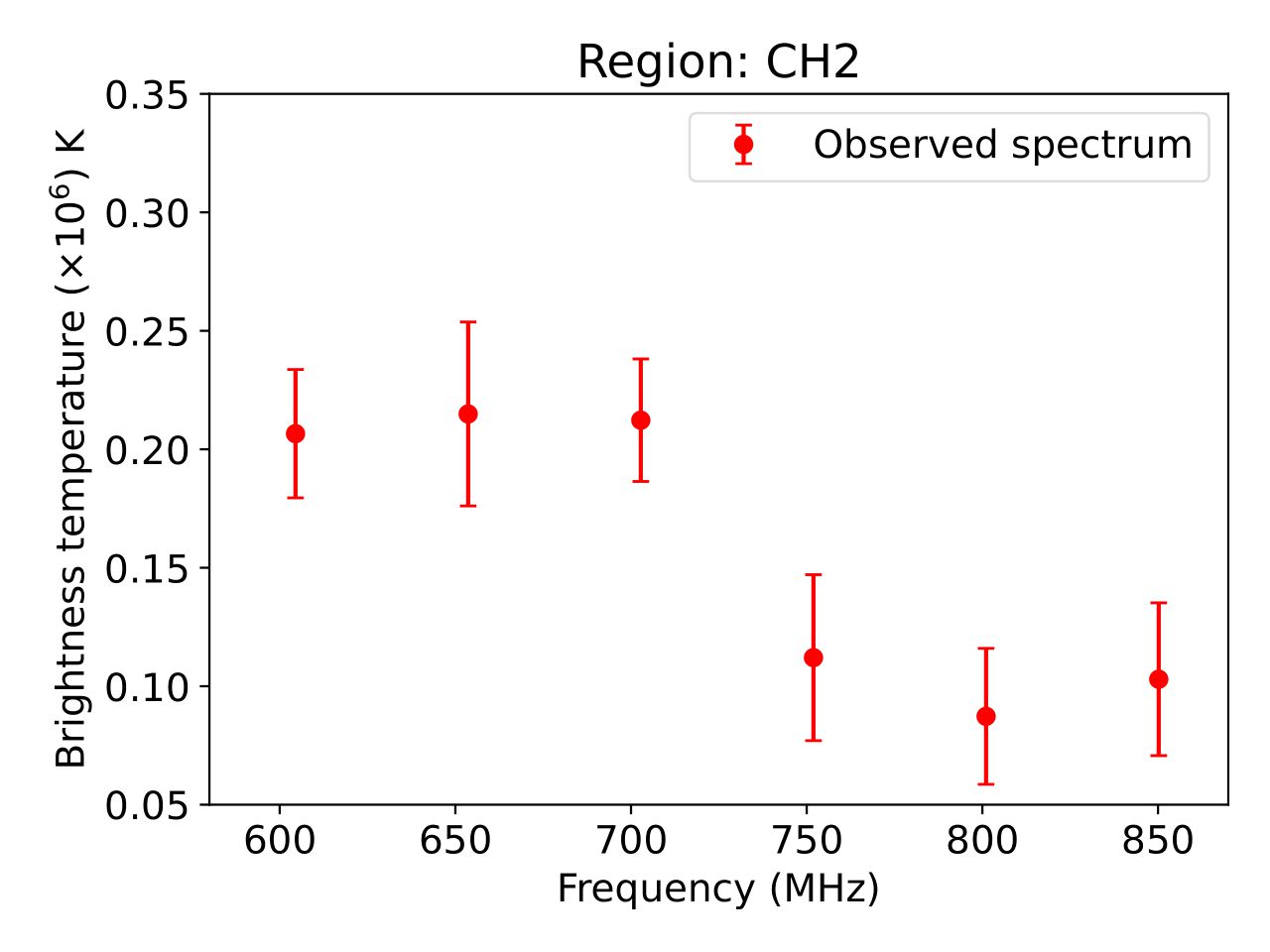}\includegraphics[trim={0cm 0cm 0cm 0cm},clip,width=0.33\linewidth]{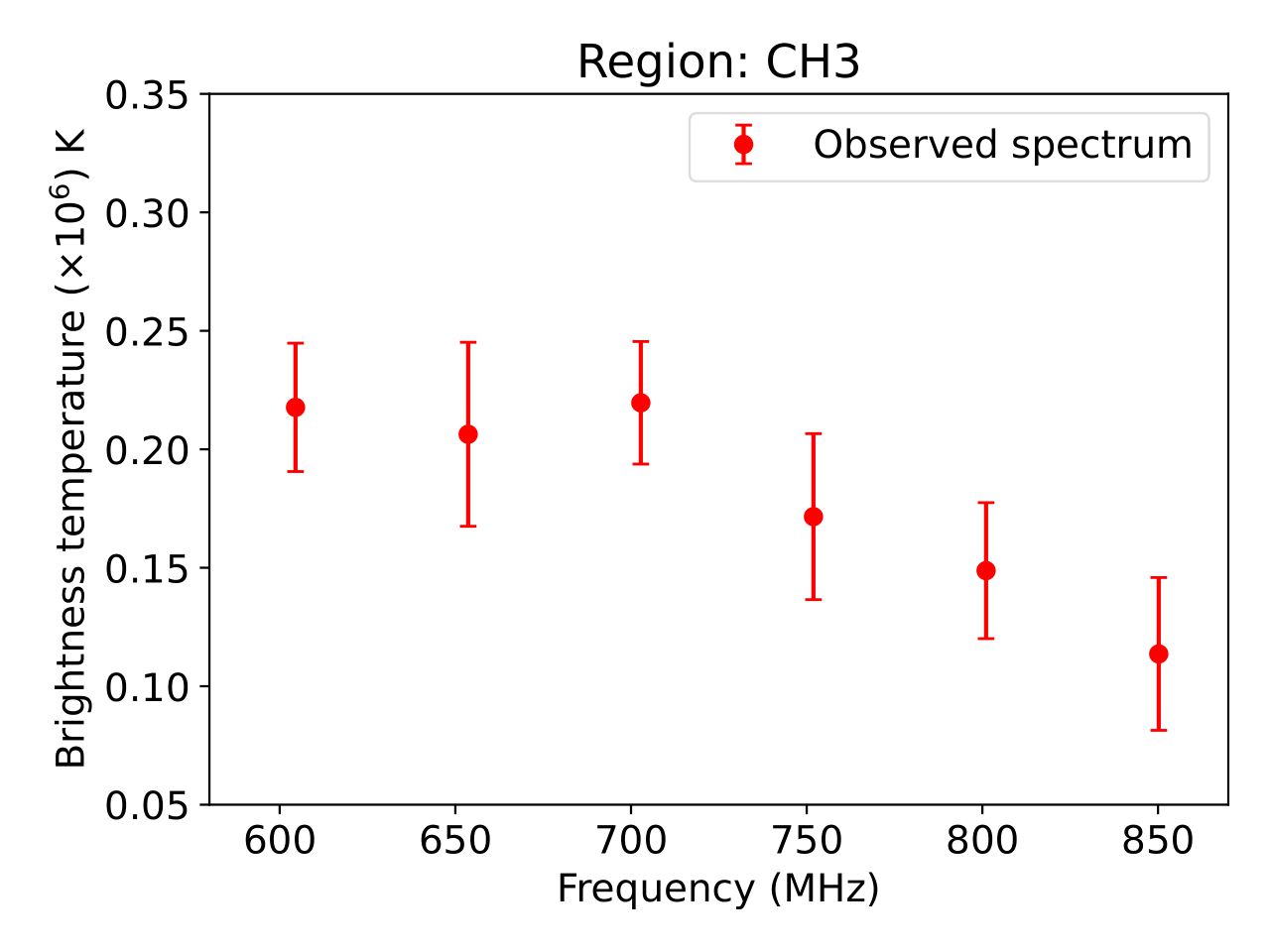}\\
\includegraphics[trim={0cm 0cm 0cm 0cm},clip,width=0.33\linewidth]{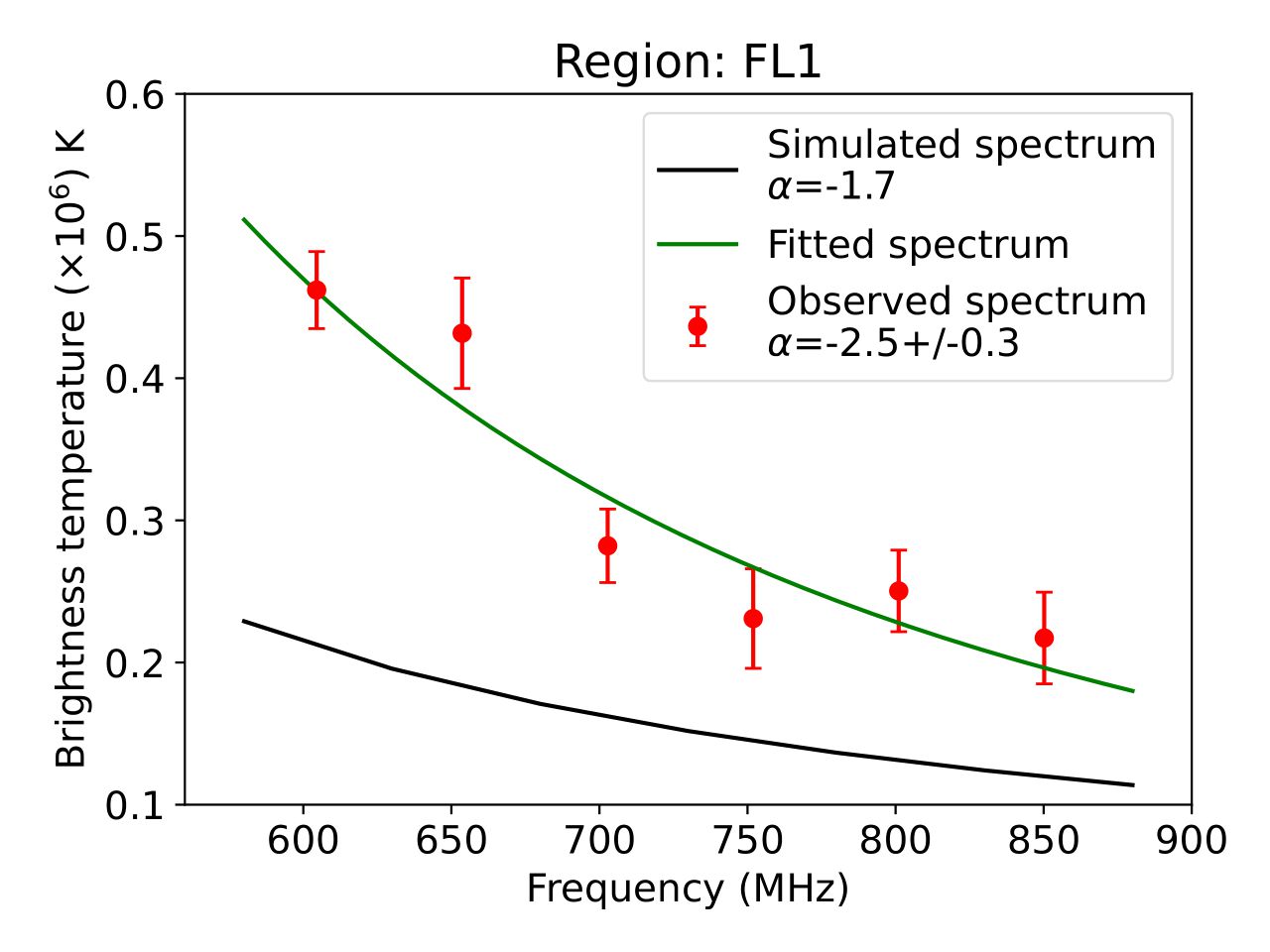}\includegraphics[trim={0cm 0cm 0cm 0cm},clip,width=0.33\linewidth]{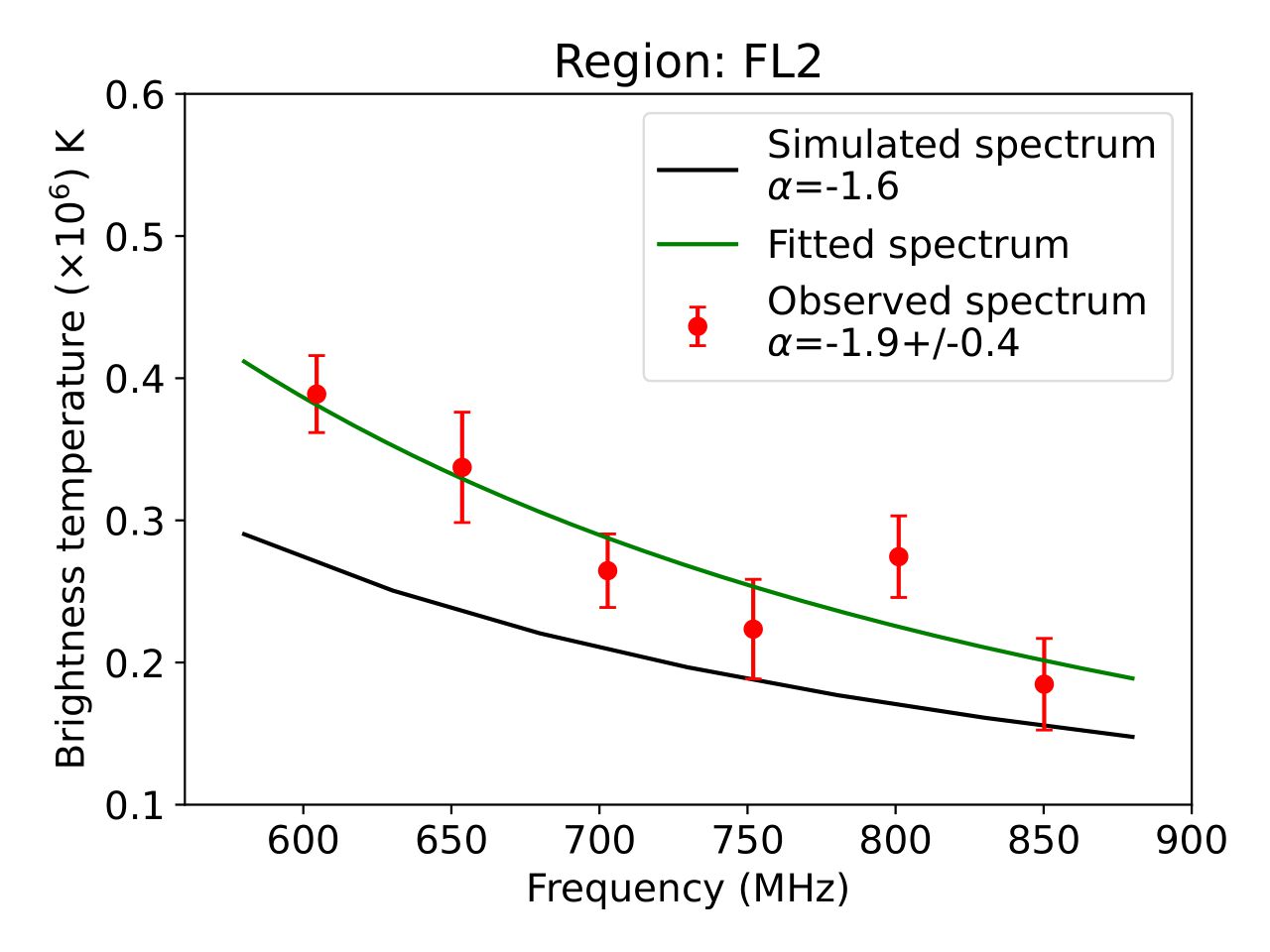}\includegraphics[trim={0cm 0cm 0cm 0cm},clip,width=0.33\linewidth]{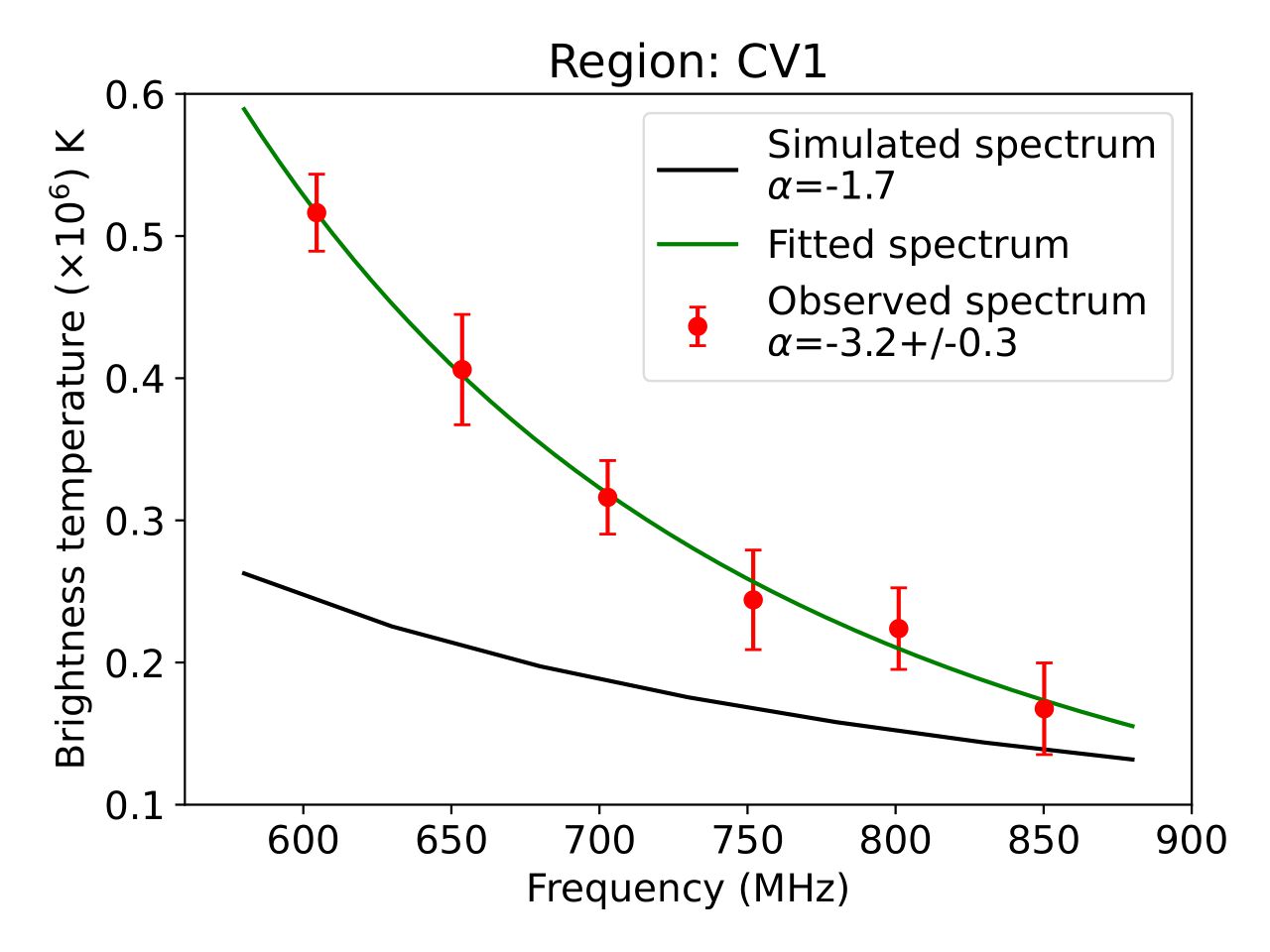}
\caption{Spatially resolved spectra from the marked PSF-sized regions in Figure \ref{fig:reg_plasma_freq}. Only spectral points with more than 3$\sigma$ detection significance are shown. The first row shows the spectra for limb regions, LR1, LR2, and LR3. The second row shows the spectra for on-disk regions, DR1, DR2, and DR3. The third row shows the spectra for coronal hole regions, CH1, CH2, and CH3. The fourth row shows the spectra from filament regions FL1 and FL2 and a coronal cavity region, CV1, associated with that filament.}
\label{fig:spectra}
\end{figure*}

Figure \ref{fig:spectra} shows spatially resolved radio spectra from the quiescent Sun regions. Each of these has been extracted from PSF-sized regions marked in the left panel of Figure \ref{fig:reg_plasma_freq}. The observed spectra (red points) are fitted with a power law, $T_\mathrm{B} \propto \nu^\alpha$ (solid green line), while simulated coronal spectra considering only free-free emission based on SDO/AIA-derived DEMs \citep{Kansabanik_2024_meerkat} are shown as solid black lines. The first row of Figure \ref{fig:spectra} shows spectra from limb regions (LR1–LR3), located at $\sim0.05\ R_\odot$ above the solar surface, where emission originates purely from coronal plasma, resulting in good agreement between observed and simulated spectra. The second row presents on-disk spectra (DR1–DR3), which come from plasma above the chromosphere and may include contributions from low-temperature TR plasma. Considering the TR optical depth model from \citet{Alissandrakis1980,Alissandrakis2020}, we have found that while free-free optical depth is close to unity in active regions \cite{Gary1989,Aschwanden2005}, for quiescent solar regions, it could be smaller than one. Hence, TR cooler plasma may contribute to on-disk decimetric MeerKAT emission in addition to coronal emission. As low-temperature plasma contributes more at lower frequencies \citep{Nindos2020,Alissandrakis2020}, spectra appear steeper than coronal-only simulations, providing a hint of the contribution from TR cool plasma not captured by EUV-based DEMs.  

Slit-based EUV spectrograph observations have shown that there is a temperature minimum of EM at log $T\sim5.2$ \citep{Raymond1981,Del2018}, below which there is a steep increase in EM toward the chromosphere. While the full-disk EM distribution maps above log $T\gtrsim5.2$ are routinely available from EUV observations from SDO/AIA, similar maps below log $T\sim5.2$ are rare. We anticipate that MeerKAT observations can provide these full-disk EM maps, including plasma at lower temperatures. However, that requires careful modeling of the TR \citep{Alissandrakis1980,Alissandrakis2020} and multi-thermal free-free radiative transfer \citep{Fleishman2021}, which is beyond the scope of this paper.

\subsubsection{Study of Coronal Holes, Filaments and Coronal Cavities}\label{subec:ch_fl_cv}
Coronal holes (CHs), characterized by open magnetic fields and reduced density and temperature \citep{Cranmer2009_coronalhole}, appear as dark regions in EUV and are key sources of fast solar wind. A prominent CH, labeled as region 7 in Figure \ref{fig:morphological_compaison}, shows spectra from three PSF-sized regions (CH1–CH3) in the third row of Figure \ref{fig:spectra}. Unlike other regions, CH spectra flatten below $\sim700$ MHz, saturate at $T_\mathrm{B} \sim 0.25$ MK, and can not be fit by a single power-law. This suggests emission from a cool, optically thick plasma layer unique to CHs. Further detailed modeling is needed to interpret these observations in detail and is beyond the scope of this work.

Coronal filaments and prominences \citep{Parenti2014} are cool, dense, elongated structures suspended in the corona along polarity inversion lines and supported by sheared magnetic fields, typically observed in H$\alpha$ and EUV absorption. Surrounding them are coronal cavities, low-density, magnetically structured voids seen as dark, circular or elliptical regions in EUV and soft X-rays, especially at the limb \citep{Gibson_2006, Fuller_G_2009}. Often associated with magnetic flux ropes and potential CME precursors, these cavities provide key pre-eruptive diagnostics \citep{Forland2013,Gibson2015}. Past radio observations using NRH \citep{Marque2004} detected such cavities and estimated densities under isothermal assumptions, but were limited by narrow spectral coverage, which is now overcome by MeerKAT. Spectra from two filament regions (FL1 and FL2) and the coronal cavity (CV1) are shown in the bottom row of the Figure \ref{fig:spectra}. The presence of cool plasma can be inferred based on a reason along the same lines as presented 
%is evident due to a similar reason mentioned 
in Section \ref{subsec:non_ar}. Although CV1 is at the limb, when compared to the %as in 
other limb spectra (LR1-LR2), the spectrum in the CV1 region shows significant steepening compared to the simulated spectrum. A forthcoming study investigating filaments and coronal cavities will explore the capability of MeerKAT for probing pre-eruptive solar phenomena.

\begin{figure*}
    \centering
    \includegraphics[trim={0cm -0.6cm 0cm 0cm},clip,width=0.44\linewidth]{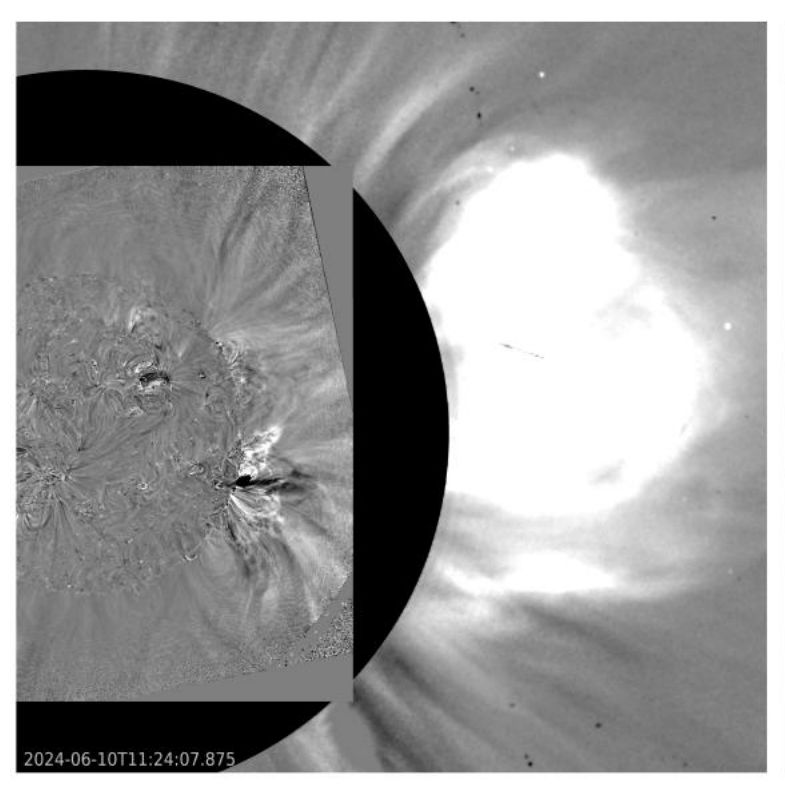}\includegraphics[width=0.5\linewidth]{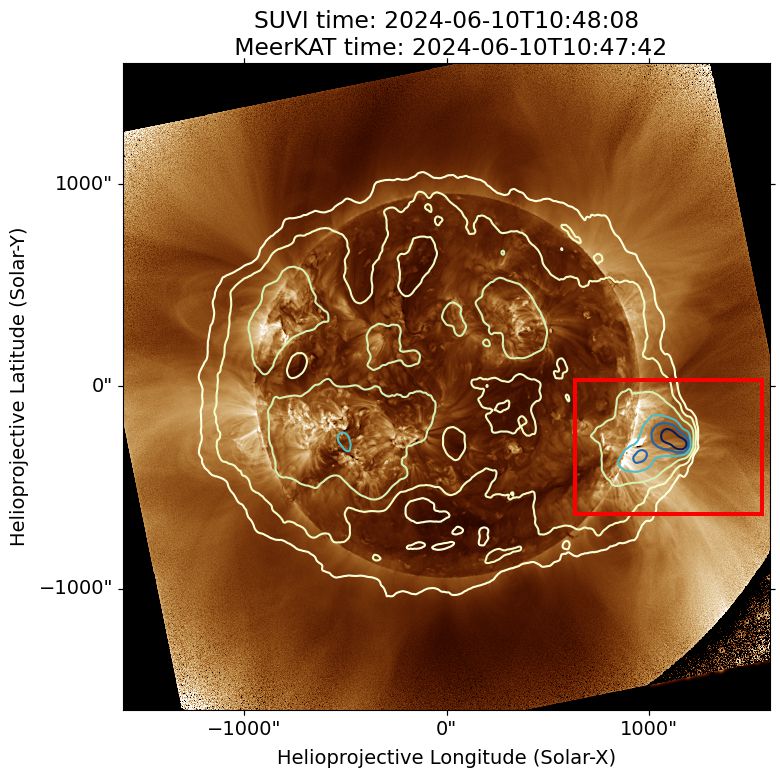}\\
    \includegraphics[width=\linewidth]{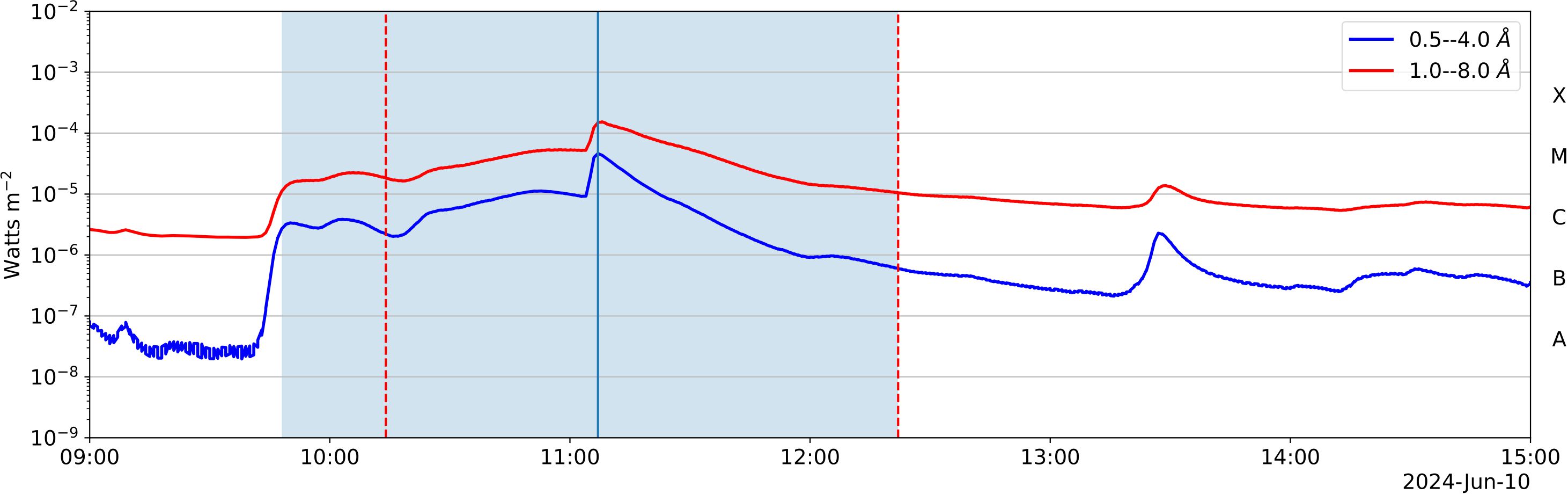}
    \caption{Flare and CME eruption on 2024 June 10. The top left panel shows base-difference composite images of the western limb eruption, with SUVI 195 \AA\ EUV images at the center and LASCO C2 coronagraph images around. In the top-right panel, MeerKAT 629 MHz radio contours at 5\%, 10\%, 20\%, 40\%, 60\%, and 80\% of the peak are overlaid on SUVI, with the red box marking the eruption site. The bottom panel shows the GOES X-ray light curve; the shaded region (09:48–12:22 UTC) indicates M-class flux levels. Red dashed lines mark the MeerKAT observation window.}
    \label{fig:cme_and_goes}
\end{figure*}

\begin{figure*}
    \centering
    \includegraphics[trim={0cm 0cm 0cm 0cm},clip,width=\linewidth]{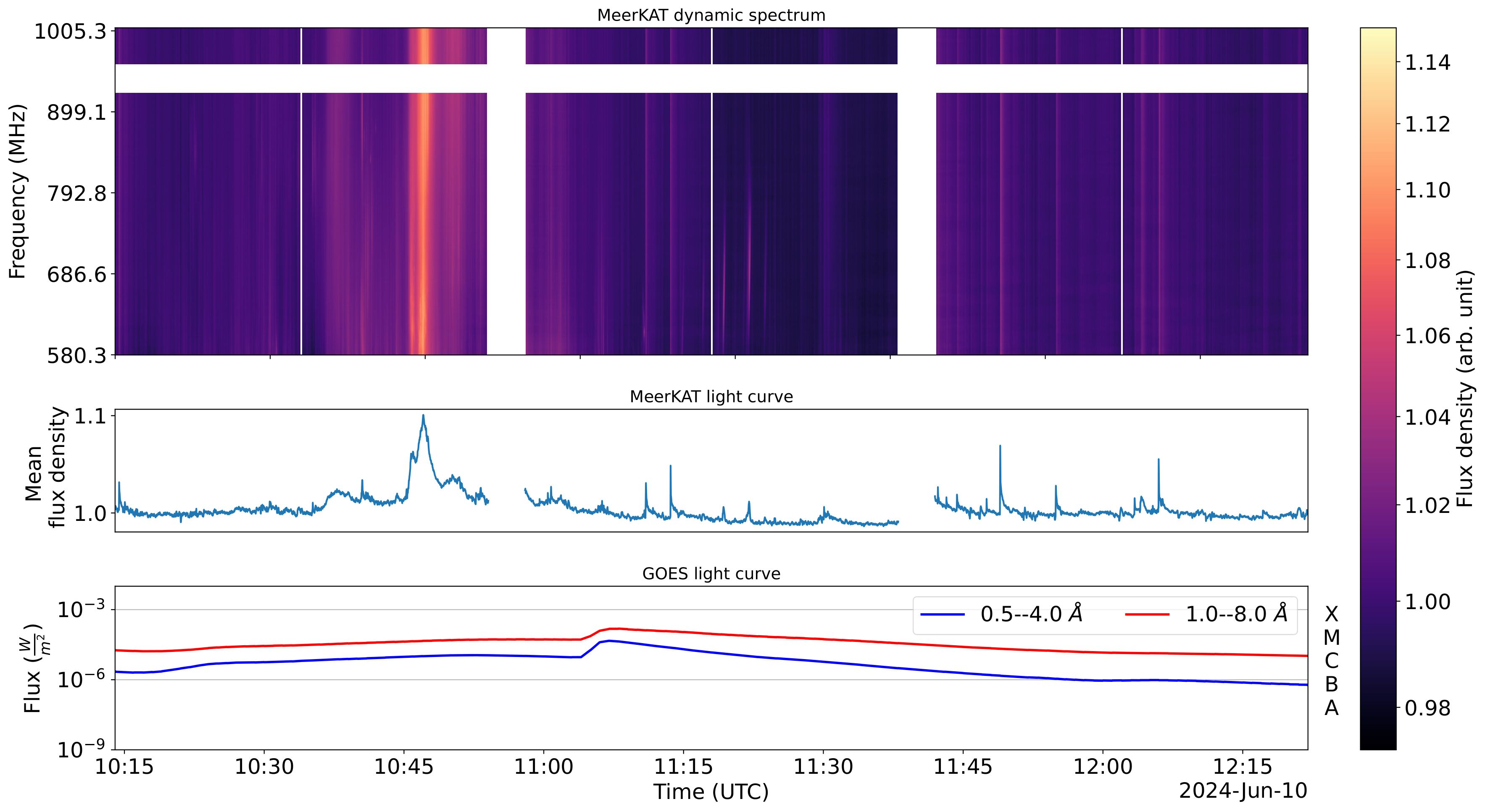}\\
    \includegraphics[trim={0cm 0cm 5cm 0.7cm},clip,width=0.49\linewidth]{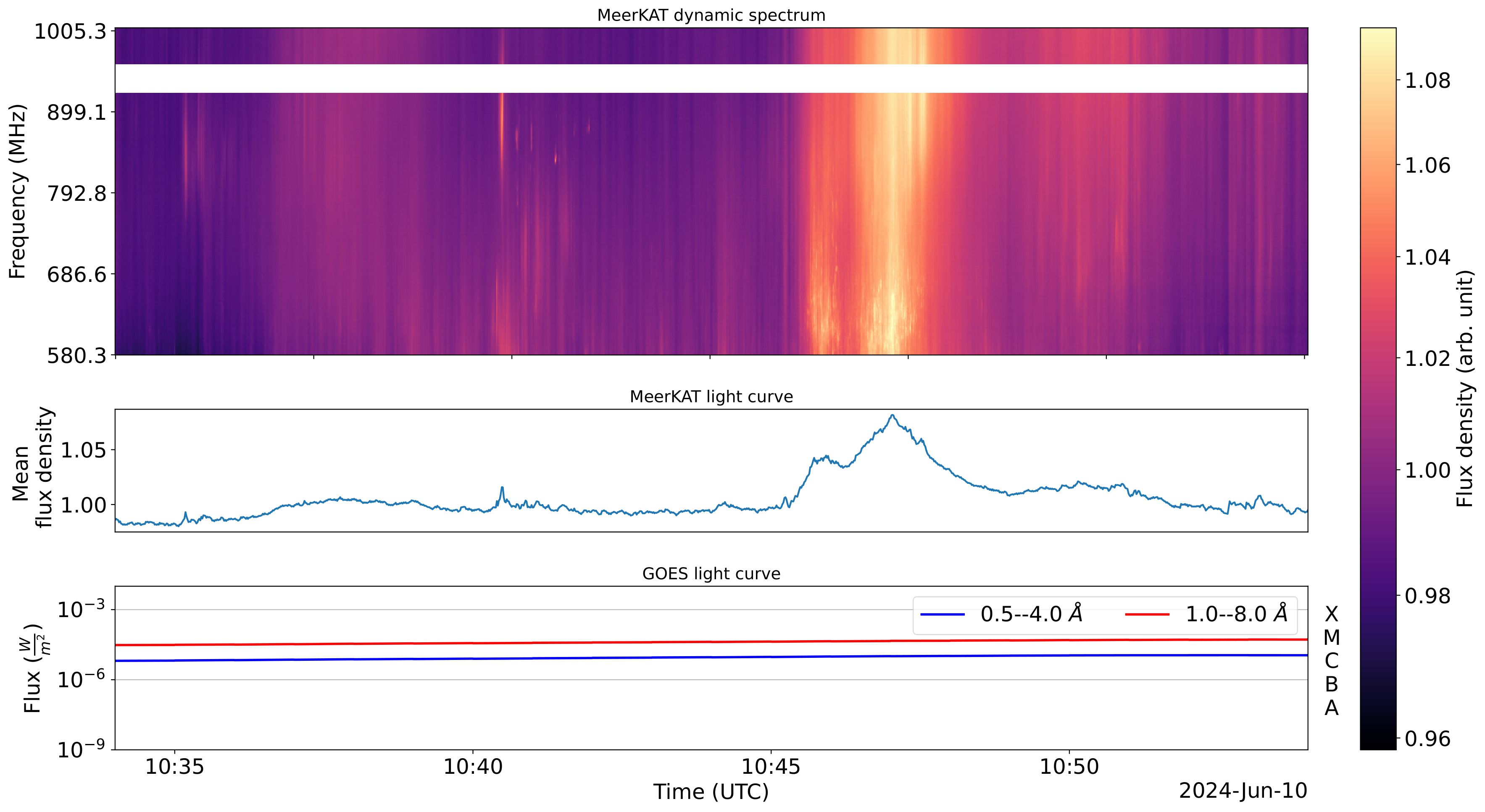}\includegraphics[trim={3.2cm 0cm 0cm 0.7cm},clip,width=0.512\linewidth]{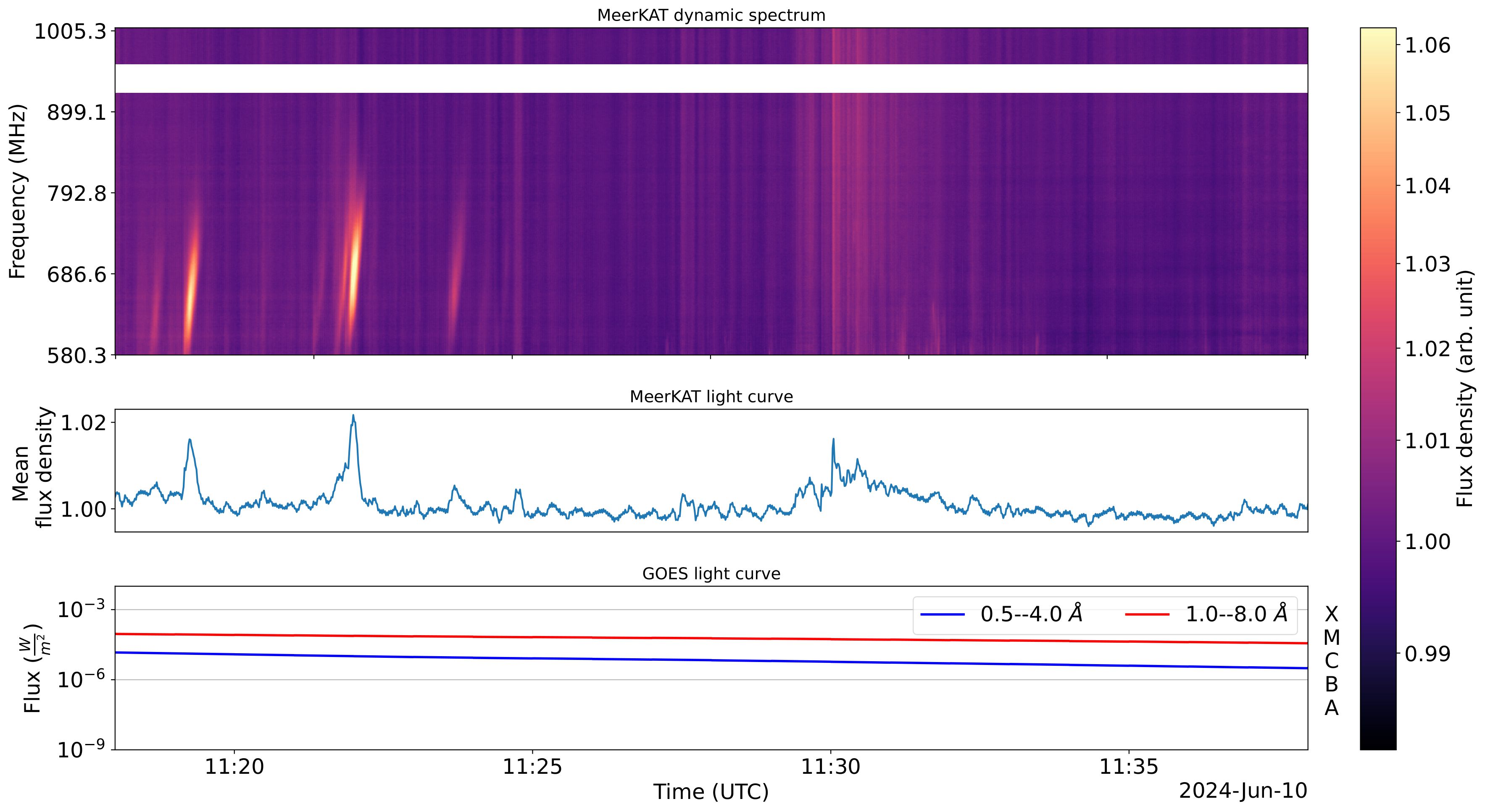}
    \caption{The top panel displays the complete duration of solar observations with MeerKAT on 2024 June 10. The bottom panels provide a zoomed-in view of the second and fourth scans. In all figures, the upper subpanels show MeerKAT total power dynamic spectrum, the middle subpanels present the band-averaged time series, and the lower subpanels plot the corresponding GOES X-ray light curve.}
    \label{fig:jun10_meerkat}
\end{figure*}

\subsection{Study of Solar Eruptions and Non-thermal Energy Release: Flares, CMEs and Associated Radio Bursts }\label{subsec:solar_eruptions}
Solar eruptions are explosive phenomena that occur in the solar atmosphere, involving the sudden release of vast amounts of energy stored in the magnetic field due to coronal dynamics. These events include solar flares, CMEs, and eruptive prominences, and are manifestations of magnetic reconnection and plasma instabilities in the solar corona. They can accelerate particles to high energies and expel large amounts of magnetized plasma in the form of CMEs into the heliosphere. These energetic particles and CMEs play a crucial role in driving space weather. Understanding the evolution of magnetic fields during their initiation, evolution, and propagation is essential for a deeper understanding of these phenomena. Spectroscopic radio imaging plays a crucial role in providing the magnetic field measurements remotely during these eruptions \citep{Vourlidas2020,Alissandrakis2021,Carley_2021} as well as providing an estimation of non-thermal particles associated with these processes. Both high-frequency ($>1$ GHz) \citep{binchen2020,gregoryfleishman2020} and low-frequency (meter-wavelength) \citep{bastian2001,Mondal2020a,Kansabanik_cme1,Kansabanik2024_cme2} observations have demonstrated their capabilities for probing non-thermal particles and measuring magnetic fields at the flare site and CME plasma at higher coronal heights, respectively. However, these eruptions lack observational probes in a crucial region in the lower corona, both in white-light, EUV, and radio wavelengths. Recently, new-generation visible light instruments, PROBA-3 \citep{proba3} and Aditya-L1/VELC \citep{velc2025}, and SunCET \citep{MasonSuncet2021} in the EUV, have been designed to observe this coronal region (at heliocentric distances $\lesssim2\ R_\odot$). The frequency range and high fidelity spectroscopic snapshot imaging capability of MeerKAT make it highly suitable for observing these eruptions between heliocentric distances of $\sim1-2\ R_\odot$.

We observed the Sun with MeerKAT from 2024 June 9–11 as part of SSV observations, targeting the active region (AR) NOAA (National Oceanic and Atmospheric Administration) AR 13711 and the reappearance of AR 13664, which was responsible for the May 2024 geomagnetic storm, anticipating continued activity. The top left panel of Figure \ref{fig:cme_and_goes} shows a composite image of a CME  event which erupted from the western limb on 10 June 2024, at 09:40 UTC, as generated using JHelioviewer \citep{helioviewer2017}, with the central image showing GOES/SUVI 195 \AA\ EUV image, and the outer panel showing SOHO/LASCO \citep{lasco1995,soho1995} C2 base-difference coronagraph image. The top right panel shows an image from MeerKAT at 629 MHz with contours overlaid on the SUVI image, with the eruption site marked by the red box. During this observing window, a long-duration X-class flare peaking at 11:07 UTC (blue line in the bottom panel of Figure \ref{fig:cme_and_goes}) occurred and remained above M-class level for a large period, from approximately 09:48 to 12:22 UTC (blue shaded region).

Several intense solar radio bursts were observed during this period in MeerKAT total power normalized dynamic spectrum, as shown in the top panel of Figure \ref{fig:jun10_meerkat}. A prominent, broadband burst around 10:46 UTC is highlighted in the bottom left panel, while several reverse-drifting bursts, typically associated with sunward-traveling electrons, are seen in the bottom right panel. Since dynamic spectra provide only spatially integrated information, distinguishing overlapping emissions requires spatially resolved dynamic spectra \citep[SPREDS;][]{Mohan2017_spreads}. Previously demonstrated at meter wavelengths using the MWA, the high-dynamic-range spectroscopic snapshot capability of MeerKAT at GHz frequencies offers similar potential. A detailed SPREDS-based analysis of these radio bursts is beyond the scope of this work.

\begin{figure*}[!htbp]
\centering
\includegraphics[width=\linewidth]{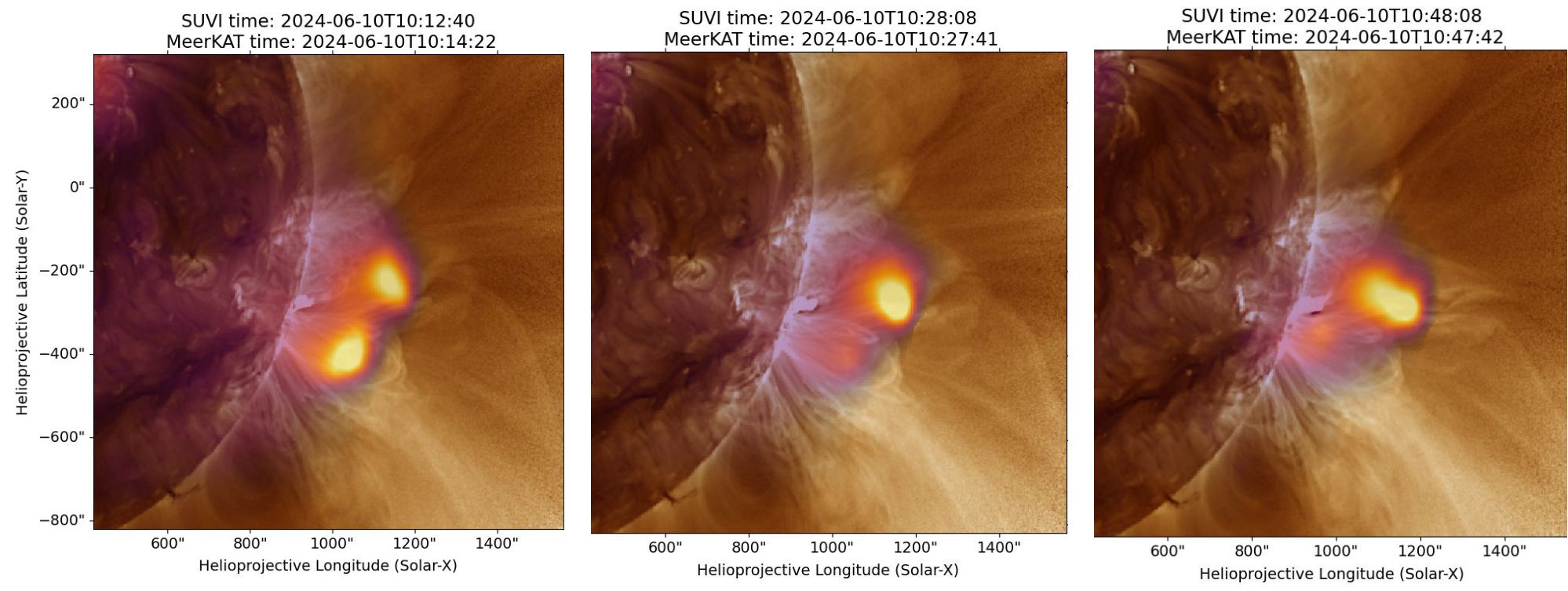}\\
\includegraphics[trim={0.5cm 0.5cm 0cm 0.2cm},clip,width=0.52\linewidth]{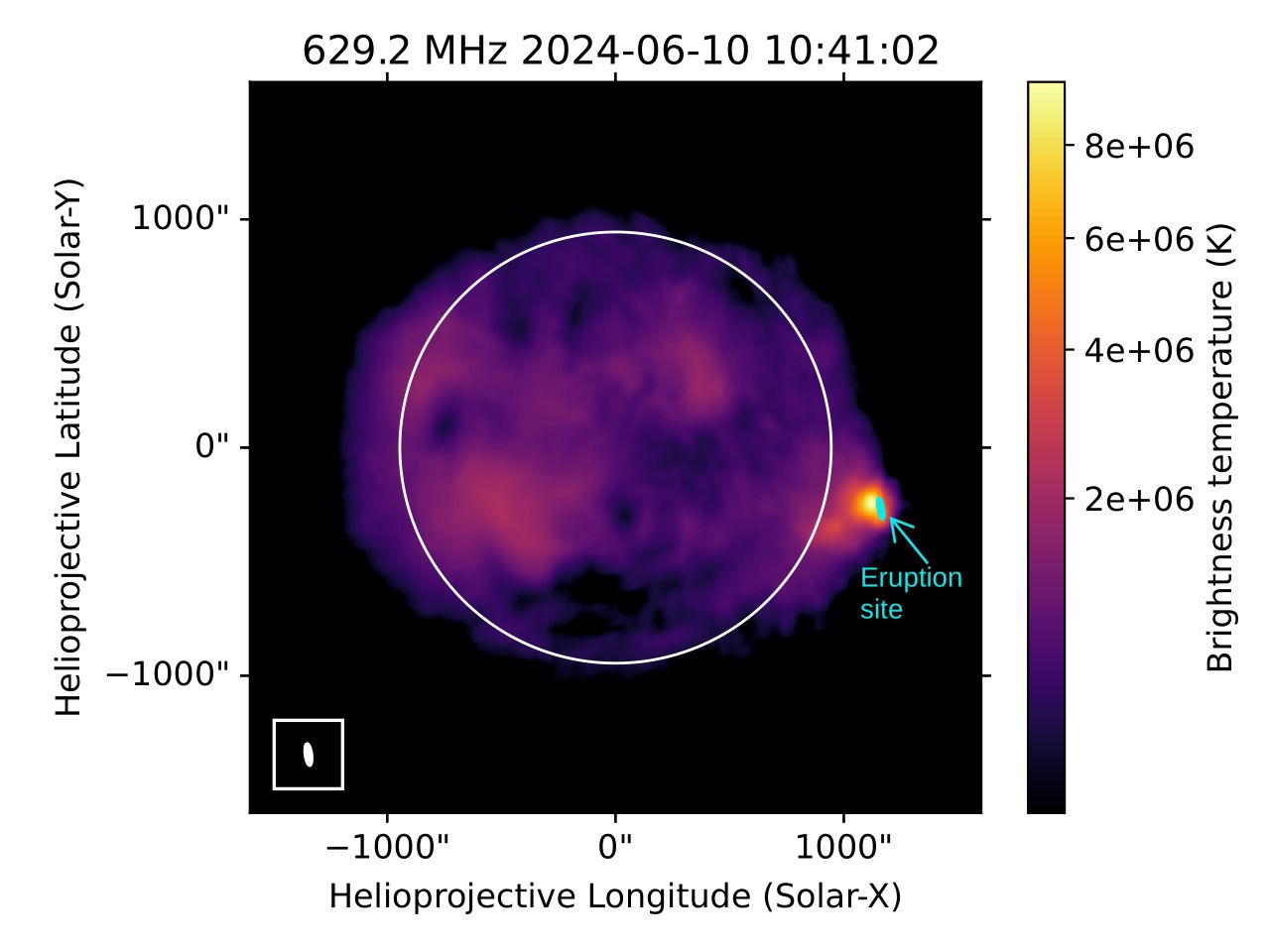}\includegraphics[trim={0.7cm 0.8cm 0.5cm 0.5cm},clip,width=0.48\linewidth]{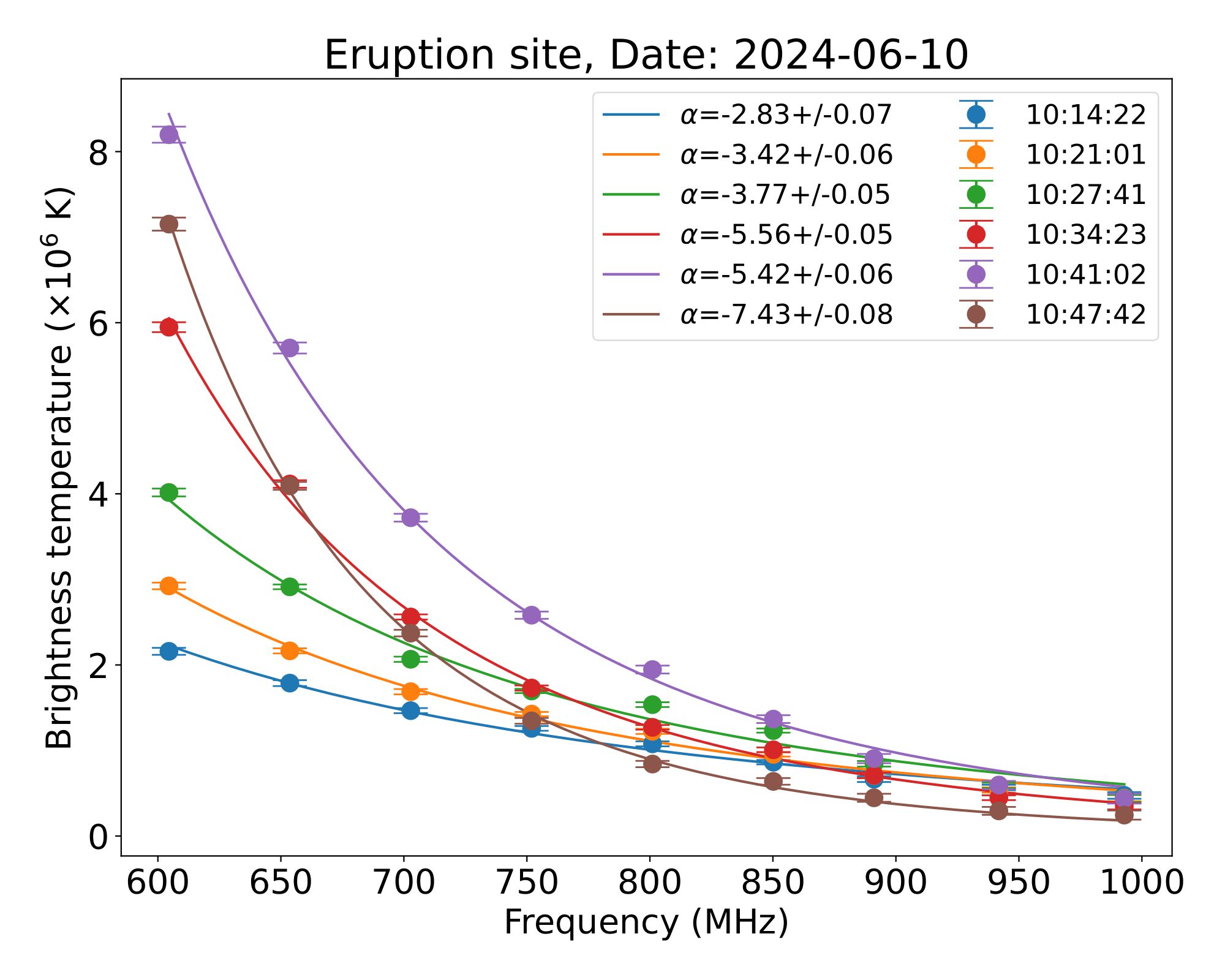}
\caption{The top panels display overlays of MeerKAT 629 MHz radio images (in a reddish-yellow colormap) on SUVI 195 \AA\ images. The bottom left panel shows the brightness temperature map at 629 MHz during the CME, highlighting a PSF-sized region. The corresponding spectra at various timestamps from this region are shown in the bottom right panel. Timestamps and spectral indices for each of the spectra are shown in the legend.}
\label{fig:cme_snapshot_spectra}
\end{figure*}

Three snapshot radio images at 629 MHz overlaid on SUVI images during the CME eruption are shown in the top panels of Figure \ref{fig:cme_snapshot_spectra}. The bottom left panel highlights the eruption site. The region from where the spectra have been extracted at multiple timestamps is marked and the spectra are shown in the bottom right panel. A progressive steepening of the spectra with time is observed, suggesting a non-thermal origin of the emission. Spectroscopic snapshot imaging at high spectral and temporal cadence can offer valuable insights into the evolving physical conditions at the eruption site, from pre-eruption to post-eruption. A detailed analysis will be presented in a future study.

\section{Conclusion and Future Works}\label{sec:conclusion}
This study has demonstrated the technical readiness of MeerKAT for carrying out well-calibrated solar observations with the telescope pointed directly at the Sun. Despite being originally designed for the observations of faint and distant galactic and extragalactic sources, MeerKAT can now be effectively utilized for solar science,  enabled by the development of a specialized observing mode and a tailored calibration and imaging pipeline. Several technical challenges associated with solar observations -- such as the need to attenuate the intense solar flux density to get the signal to lie within the linear range of the receivers, perform reliable flux density calibration using internal noise diodes, and account for the sidereal drift of the Sun during the observation -- have been addressed. The resulting calibrated images show strong morphological agreement with EUV observations and expectations based on simulations, validating the viability of using MeerKAT for solar radio imaging and also demonstrating the robustness of the interferometric calibration and imaging pipeline. 
We note that MeerKAT solar observations at the upper end of the L-band and the S-band will be affected by missing flux density issues due to the limited availability of sufficiently short baselines.
Potential solutions to this issue will involve incorporating single-dish total power measurements from MeerKAT itself, or from other instruments, into the interferometric imaging process.

Nonetheless, these developments offer access to a previously underexplored observational regime, enabling high spatial and spectro-temporal resolution studies of the solar atmosphere. Even the limited observations obtained during this work highlight the valuable novel insights that can be gained into the structure and dynamics of the quiet corona, weak transient events, and large-scale eruptive phenomena such as flares and CMEs. Additionally, the polarimetric capabilities of MeerKAT hold promise for coronal magnetic field studies when combined with multi-wavelength space-based and in-situ data. Ongoing efforts to characterize and validate its polarization response for solar observations aim to support such advancements and pave the way for future breakthroughs. We hope that the work presented here will also provide motivation and guidance for commissioning the solar observing mode alongside other observing modes for the upcoming next-generation instruments like the SKA-mid \citep{Plunkett_2023}.

\section{Appendix}
\subsection{Choice of Frequency Dependent Multiscale Parameters}\label{app:multiscale_params}
Multiscale deconvolution has two primary parameters -- multiscale-scales and multiscale-bias. Multiscale-scales define the angular sizes used in multiscale deconvolution to separate emission structures of different angular extents, while the multiscale-bias parameter controls the preference for selecting smaller or larger scales during the cleaning process. The largest multiscale-scale is set to larger of half of the solar disc size and the maximum recoverable angular scale of the array, $S_\mathrm{max}$. Additional multiscale-scales include the image pixel scale, number of pixels inside the FWHM of the PSF ($w$), and steps in multiples of 2 till $S_\mathrm{max}$. This choice effectively captures both compact and diffuse coronal features. 

In a radio interferometric array, a baseline vector $\vec{b} = (b_\mathrm{x}, b_\mathrm{y})$ corresponds to a {\it uv}-coordinate at observing frequency $\nu$ is given by,
\begin{equation}
   (u, v) = \frac{\nu}{c}\ (b_\mathrm{x}, b_\mathrm{y}),
\end{equation}
where $c$ is the speed of light. Each spectral channel samples a slightly different frequency, and thus contributes a slightly different {\it uv}-point. The spread in {\it uv}-coordinates across the bandwidth, $\delta \nu$, is,
\begin{equation}
   \Delta u \sim \frac{\delta \nu}{c}\ b_\mathrm{x}, \quad \Delta v \sim \frac{\delta \nu}{c}\ b_\mathrm{y}. 
\end{equation}
The area in the {\it uv}-plane covered by a single baseline due to finite bandwidth is therefore proportional to
\begin{equation}
    \Delta u \Delta v \propto \left( \frac{\delta \nu}{\nu} \right)^2 (u^2 + v^2).
\end{equation}
Hence, the local {\it uv}-coverage density $\rho_{\mathrm{uv}}$ scales as,
\begin{equation}
    \rho_{\mathrm{uv}} \propto \left( \frac{1}{\delta \nu / \nu} \right)^2.
\end{equation}

Since the fractional bandwidth $\delta \nu / \nu$ is larger at lower frequencies for a fixed channel width $\delta \nu$, the same array configuration provides denser {\it uv}-coverage in the UHF band than in the L-band. The sparseness ($1/\rho_\mathrm{uv}$) is inversely proportional to frequency, hence in log-space, it decreases linearly. We found that in the UHF band, the default multiscale bias parameter in \textsf{WSClean} \citep{offringa2017_wsclean_multiscale} works optimally. However, at the higher frequencies within the L-band, where sensitivity to large-scale structures diminishes as well as the sparseness in {\it uv}-coverage increases, the default bias tends to introduce spurious large-scale emissions. To mitigate this, we introduce a frequency-dependent multiscale-bias parameter that reduces bias toward large angular scales linearly in log-space. This is defined as:
\begin{equation}
\begin{split}
\text{b}(\nu) = 
\begin{cases}
    b_{\min}, \text{if } \nu \leq 1015\ \text{MHz} \\
    b_{\max}, \text{if } \nu \geq 1670\ \text{MHz} \\ 
    b_{\min} + \left( \dfrac{\log \nu - \log \nu_{\min}}{\log \nu_{\max} - \log \nu_{\min}} \right) 
    (b_{\max} - b_{\min}), \\ \text{otherwise}.
\end{cases}
\end{split}
\label{eq:bias_parameter}
\end{equation}
where, $b_{\min}$ and $b_{\max}$ are chosen as 0.6 and 0.9, respectively. This formulation progressively increases the bias toward smaller scales at higher frequencies, while retaining sensitivity to extended emission at lower L-band frequencies.

\section*{Acknowledgments}
This work uses observations from MeerKAT radio telescope. MeerKAT telescope is operated by the South African Radio Astronomy Observatory, which is a facility of the National Research Foundation, an agency of the Department of Science, Technology, and Innovation. The authors acknowledge the contribution of all those who designed and built MeerKAT instrument. D. K. acknowledges support for this research by the NASA Living with a Star Jack Eddy Postdoctoral Fellowship Program, administered by UCAR’s Cooperative Programs for the Advancement of Earth System Science (CPAESS) under award 80NSSC22M0097. D.P. and D.O. acknowledge support from the Department of Atomic Energy, Government of India, under the project no. 12-R\&D-TFR-5.02-0700. A.V. was supported by NASA grants 80NSSC21K1860 and 80NSSC22K0970.  D.K. gratefully acknowledges Justin Jonas and Thomas Abbott (SARAO) for their valuable feedback on the early draft of the manuscript. D.K. gratefully acknowledges Surajit Mondal (NJIT, USA) for useful discussions and suggestions. We also gratefully acknowledge the thoughtful and positive comments from reviewers, which have helped to improve the clarity and presentation of this work.

\bibliography{manuscript}{}
\bibliographystyle{Frontiers-Harvard}
\end{document}